\documentclass[11pt]{article}
\usepackage{amsfonts,latexsym,graphicx,url}
\usepackage{amsthm,amsmath,amsfonts,graphicx}
\usepackage[sort&compress,numbers]{natbib}
\title{The Distance Geometry of Music}

\author{%
  Erik D. Demaine%
    \thanks{Computer Science and Artificial Intelligence Laboratory,
      Massachusetts Institute of Technology, Cambridge, Massachusetts, USA, 
      \protect\url{edemaine@mit.edu}}
\and
  Francisco Gomez-Martin%
    \thanks{Departament de Matem\'atica Aplicada,
      Universidad Polit\'ecnica de Madrid, Madrid, Spain,
      \protect\url{fmartin@eui.upm.es}}
\and
  Henk Meijer%
    \thanks{School of Computing, Queen's University, Kingston, Ontario, Canada,
      \protect\url{{henk,daver}@cs.queensu.ca}}
\and
  David Rappaport\footnotemark[3]
\and
  Perouz Taslakian%
    \thanks{School of Computer Science, McGill University, Montr\'eal,
      Qu\'ebec, Canada, \protect\url{{perouz,godfried}@cs.mcgill.ca}}
\and
  Godfried T. Toussaint\footnotemark[4]
  \thanks{Centre for Interdisciplinary Research in Music Media and Technology
          The Schulich School of Music McGill University.
          Supported by FQRNT and NSERC.}
\and
  Terry Winograd%
    \thanks{Department of Computer Science, Stanford University, Stanford, California, USA,
      \protect\url{winograd@cs.stanford.edu}}
\and
  David R. Wood%
    \thanks{Departament de Matem\`atica Aplicada II,
      Universitat Polit\`ecnica de Catalunya, Barcelona, Spain,
      \protect\url{david.wood@upc.edu}.
      Supported by the Government of Spain grant MEC SB2003-0270, 
      and by the projects MCYT-FEDER BFM2003-00368 and Gen.\ Cat 2001SGR00224.}
}

\date{}

\newif\ifabstract
\abstracttrue
\abstractfalse
\newif\iffull
\ifabstract \fullfalse \else \fulltrue \fi

\advance \topmargin by -\headheight
\advance \topmargin by -\headsep
\evensidemargin \oddsidemargin
\let\latexref=\ref
\def\ref{\nolinebreak\latexref}

{\makeatletter
 \gdef\xxxmark{%
   \expandafter\ifx\csname @mpargs\endcsname\relax 
     \expandafter\ifx\csname @captype\endcsname\relax 
       \marginpar{xxx}
     \else
       xxx 
     \fi
   \else
     xxx 
   \fi}
 \gdef\xxx{\@ifnextchar[\xxx@lab\xxx@nolab}
 \long\gdef\xxx@lab[#1]#2{{\bf [\xxxmark #2 ---{\sc #1}]}}
 \long\gdef\xxx@nolab#1{{\bf [\xxxmark #1]}}
 \long\gdef\xxx@lab[#1]#2{}\long\gdef\xxx@nolab#1{}%
}

\newcommand{\ceil}[1]{\ensuremath{\protect\lceil#1\rceil}}
\newcommand{\CEIL}[1]{\ensuremath{\protect\left\lceil#1\right\rceil}}
\newcommand{\FLOOR}[1]{\ensuremath{\protect\left\lfloor#1\right\rfloor}}
\newcommand{\floor}[1]{\ensuremath{\protect\lfloor#1\rfloor}}
\newcommand{\bracket}[1]{\ensuremath{\protect\left(#1\right)}}
\newcommand{\half}{\ensuremath{\protect\tfrac{1}{2}}}

\newcommand{\AAA}{\textup{(A)}}
\newcommand{\BBB}{\textup{(B)}}
\newcommand{\CCC}{\textup{(C)}}
\newcommand{\DDD}{\textup{(D)}}
\newcommand{\STAR}{\textup{$(\star)$}}
\newcommand{\STARSTAR}{\textup{$(\star\star)$}}

\theoremstyle{plain}
\newtheorem{theorem}{Theorem}[section]
\newtheorem{lemma}[theorem]{Lemma}
\newtheorem{corollary}[theorem]{Corollary}

\newtheorem{fact}[theorem]{Fact}

\theoremstyle{definition}

\newtheorem{observation}{Observation}

\def\captionfont{\small\rm}
\def\captionlabelfont{\small\bf}
{\makeatletter
 \global\let\plainfont@makecaption=\@makecaption
 \long\gdef\@makecaption#1#2{%
   \plainfont@makecaption{\captionlabelfont #1}{\captionfont #2}}}

\def\pmod#1{\allowbreak\ ({\rm mod}\,\,#1)}

\newlength\rboxWidth
\DeclareUrlCommand\rbox{%
  \gdef\ShouldAllowBreak{}%
  \def\UrlSpecials{%
    \do\x{\ShouldAllowBreak\gdef\ShouldAllowBreak{\allowbreak}\times}%
    \do\.{\hbox to \rboxWidth{\hss$\cdot$\hss}}%
  }%
  \catcode`x=11%
  \catcode`.=11%
}

\DeclareFontFamily{U}{mathbarr}{\hyphenchar\font45}
\DeclareFontShape{U}{mathbarr}{m}{n}{<-> gen * mathbarr}{}
\DeclareSymbolFont{mathbarr}{U}{mathbarr}{m}{n}
\DeclareFontSubstitution{U}{mathbarr}{m}{n}
\DeclareMathSymbol{\curvearrowleft}{3}{mathbarr}{"F0}
\DeclareMathSymbol{\curvearrowright}{3}{mathbarr}{"F1}
\DeclareMathSymbol{\curvearrowleftright}{3}{mathbarr}{"F2}

\def\clockwised{\vbox{\hbox to 0pt{$\scriptstyle\curvearrowright\hss$}\vspace{-1.4ex}\hbox{$d$}}}
\def\geodesicd{\vbox{\hbox to 0pt{$\scriptstyle\curvearrowleftright\hss$}\vspace{-1.4ex}\hbox{$d$}}}
\def\chordd{\overline{d}}

\newlength\abovealgorithmskip
\abovealgorithmskip=\bigskipamount
\newlength\midalgorithmskip
\midalgorithmskip=\smallskipamount
\newlength\belowalgorithmskip
\belowalgorithmskip=\bigskipamount
\newcommand\algorithmfont{\normalsize}
\newcommand\algorithmwordfont{\bf}
\newcommand\algorithmtitlefont{\sc}

\newbox\algorithmboxbox
\newlength\algorithmmargin
\newenvironment{algorithmbox}[1]
  {\setbox\algorithmboxbox\vbox\bgroup
     \advance\linewidth by -2\algorithmmargin
     \hsize=\linewidth
     \noindent{\algorithmwordfont Algorithm }{\algorithmtitlefont #1}\par
     \vspace\midalgorithmskip
     \algorithmfont}
  {\egroup
   \par\vspace\abovealgorithmskip
   \centerline{\fbox{\box\algorithmboxbox}}%
   \par\vspace\belowalgorithmskip}

\def\compactify{\itemsep=0pt \topsep=0pt \partopsep=0pt \parsep=0pt}
\let\latexusecounter=\usecounter
\newenvironment{itemize*}
  {\def\usecounter{\compactify\latexusecounter}
   \begin{itemize}}
  {\end{itemize}\let\usecounter=\latexusecounter}
\newenvironment{enumerate*}
  {\def\usecounter{\compactify\latexusecounter}
   \begin{enumerate}}
  {\end{enumerate}\let\usecounter=\latexusecounter}

\begin{document}
\maketitle

\begin{abstract}

We demonstrate relationships between the classic Euclidean algorithm
and many other fields of study, particularly in the context of music
and distance geometry.
Specifically, we show how the structure of the Euclidean algorithm
defines a family of rhythms
which encompass over forty timelines (\emph{ostinatos})
from traditional world music.
We prove that these \emph{Euclidean rhythms} have the mathematical property
that their onset patterns are distributed as evenly as possible:
they maximize the sum of the Euclidean distances between all pairs of onsets,
viewing onsets as points on a circle.
Indeed, Euclidean rhythms are the unique rhythms that maximize this notion
of \emph{evenness}.
%
We also show that essentially all Euclidean rhythms are \emph{deep}:
each distinct distance between onsets occurs with a unique multiplicity,
and these multiplicies form an interval $1, 2, \dots, k-1$.
Finally, we characterize all deep rhythms, showing that they form a
subclass of generated rhythms, which in turn proves a useful property called
shelling.
All of our results for musical rhythms apply equally well to musical scales.
In addition, many of the problems we explore are interesting in their
own right as distance geometry problems on the circle; some of the same
problems were explored by Erd\H{o}s in the plane.
\end{abstract}

\newpage

\section{Introduction}

Polygons on a circular lattice,
African bell rhythms~\cite{toussaint-03},
musical scales~\cite{clough-99b},
spallation neutron source
accelerators in nuclear physics~\cite{bjorklund-03a},
linear sequences in mathematics~\cite{lunnon-92},
mechanical words and
stringology in computer science~\cite{lothaire-02},
drawing digital straight lines in computer graphics~\cite{klette-04},
calculating leap years in calendar design~\cite{harris-04, ascher-02},
and an ancient algorithm for computing the greatest common divisor
of two numbers, originally described by
Euclid~\cite{euclid-56,franklin-56}---what do these disparate concepts
all have in common?
The short answer is, ``patterns distributed as evenly as possible''.
For the long answer, please read on.

Mathematics and music have been intimately intertwined
since over 2,500 years ago when the famous Greek mathematician,
Pythagoras of Samos (circa 500~B.C.),
discovered that the pleasing experience of musical
harmony is the result of ratios of small integers~\cite{ashton-03}.
Most of this interaction between the two fields, however, has been in the
domain of pitch and scales.
For some historical snapshots of this interaction, we refer the
reader to H. S. M. Coxeter's delightful account~\cite{coxeter-62}.
In music theory, much attention has been devoted to the study
of intervals used in pitch scales~\cite{forte-73}, but relatively little
work has been devoted to the analysis of time duration intervals of rhythm.
Some notable recent exceptions are the books by Simha Arom~\cite{arom-91},
Justin London~\cite{london-04} and Christopher Hasty~\cite{hasty-97}.

In this paper, we study various mathematical properties of musical rhythms
and scales that are all, at some level, connected to an algorithm of another
famous ancient Greek mathematician, Euclid of Alexandria (circa 300~B.C.).
We begin (in Section~\ref{sec:Euclid}) by showing several mathematical
connections between musical rhythms and scales, the work of Euclid,
and other areas of knowledge such as nuclear physics, calendar design,
mathematical sequences, and computer science.
In particular, we define the notion of \emph{Euclidean rhythms},
generated by an algorithm similar to Euclid's.
Then, in the more technical part of the paper
(Sections \ref{first technical section}--\ref{last technical section}),
we study two important properties of rhythms and scales,
called \emph{evenness} and \emph{deepness},
and show how these properties relate to the work of Euclid.

The Euclidean algorithm has been connected to music theory previously
by Viggo Brun~\cite{brun-64}.
Brun used Euclidean algorithms to calculate the lengths
of strings in musical instruments between two lengths $l$ and~$2l$,
so that all pairs of adjacent strings have the same length ratios.
In contrast, we relate the Euclidean algorithm to rhythms
and scales in world music.


Musical rhythms and scales can both be seen as two-way
infinite binary sequences~\cite{toussaint-02}.
In a rhythm, each bit represents one unit of time called a \emph{pulse}
(for example, the length of a sixteenth note),
a one bit represents a played note or \emph{onset}
(for example, a sixteenth note), and
a zero bit represents a silence (for example, a sixteenth rest).
In a scale, each bit represents a pitch
(spaced uniformly in log-frequency space),
and zero or one represents whether the pitch is absent or present
in the scale.
Here we suppose that all time intervals between onsets in a rhythm
are multiples of a fixed time unit, and that all tone intervals
between pitches in a scale are multiples of a fixed tonal unit
(in logarithm of frequency).  

The time dimension of rhythms and the pitch dimension of scales
have an intrinsically cyclic nature, cycling every measure
and every octave, respectively.
In this paper, we consider rhythms and scales that match this cyclic
nature of the underlying space.
In the case of rhythms, such cyclic rhythms are also called \emph{timelines},
rhythmic phrases or patterns that are repeated throughout a piece;
in the remainder of the paper, we use the term ``rhythm'' to mean ``timeline''.
The infinite bit sequence representation of a cyclic rhythm or scale
is just a cyclic repetition of some $n$-bit string, corresponding to
the timespan of a single measure or the log-frequency span of a single octave.
%
To properly represent the cyclic nature of this string, we imagine assigning
the bits to $n$ points equally spaced around a circle of
circumference~$n$~\cite{mccartin-98}.
A rhythm or scale can therefore be represented as a subset of these $n$ points.
We use $k$ to denote the size of this subset;
that is, $k$ is the number of onsets in a rhythm or pitches in a scale.
For uniformity, the terminology in the remainder of this paper speaks
primarily about rhythms, but the notions and results apply equally well
to scales.

\begin{figure}
  \centering
  \includegraphics[scale=0.6]{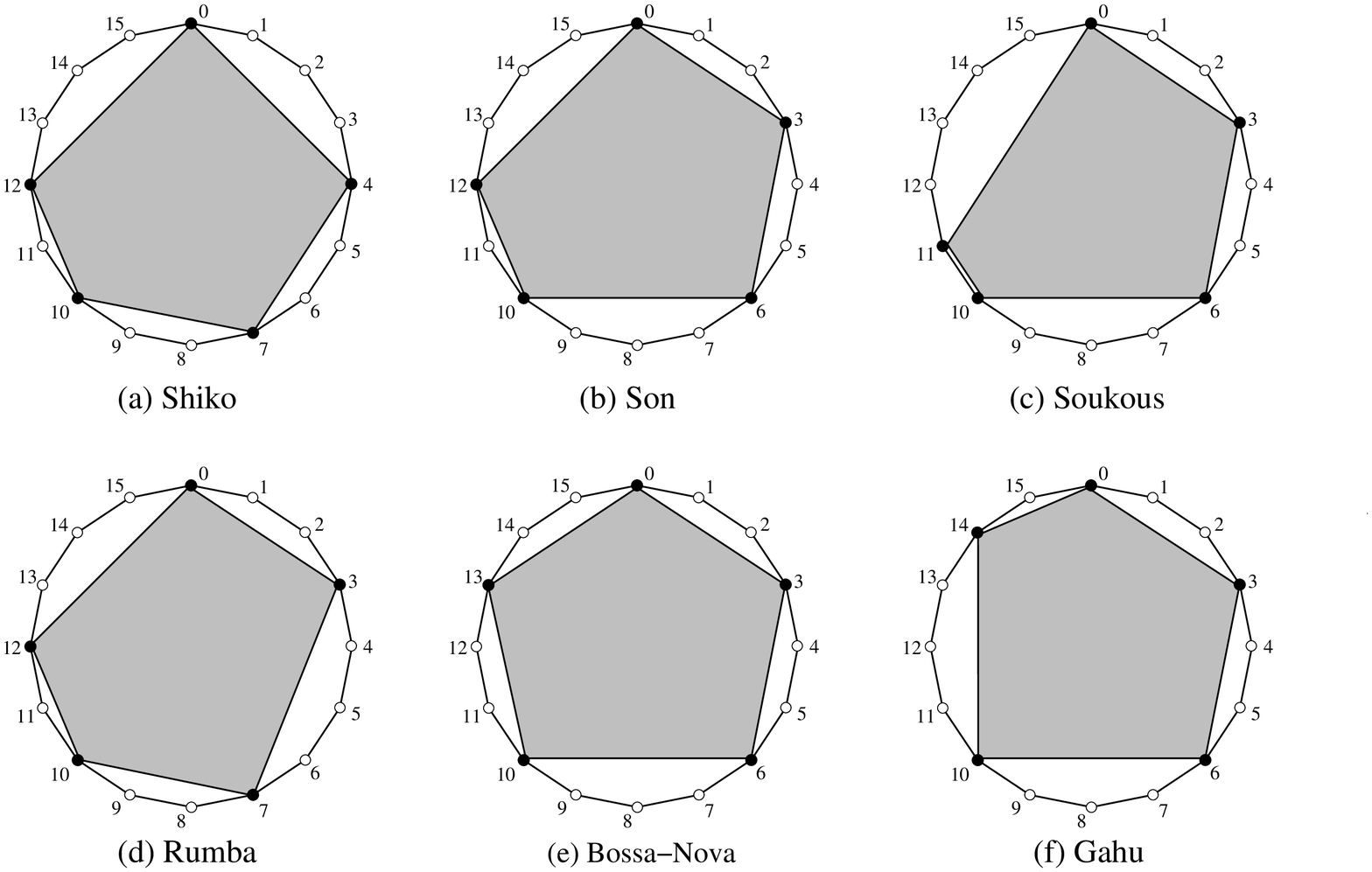}
  \caption{The six fundamental African and Latin American rhythms 
           which all have equal sum of pairwise geodesic distances; 
           yet intuitively, the Bossa-Nova rhythm is more ``even'' than the rest.} 
  \label{even-cuban}
\end{figure}

In this paper, we use four representations of rhythms of timespan~$n$. 
The first representation is the commonly used \emph{box-like} representation,
also known as the Time Unit Box System (TUBS),
which is a sequence of $n$ `\rbox{x}'s and `\rbox{.}'s
where `\rbox{x}' represents an onset
and `\rbox{.}' denotes a silence
(a zero bit)~\cite{toussaint-02}. 
This notation was used and taught in the West by Philip Harland 
at the University of California, Los Angeles, in 1962, and
it was made popular in the field of ethnomusicology by 
James Koetting~\cite{koetting-70}.  However, such box notation 
has been used in Korea for hundreds of years~\cite{hye-ku-81}.
%
The second representation of rhythms and scales we use
is the \emph{clockwise distance sequence}, 
which is a sequence of integers that sum up to $n$ and represent 
the lengths of the intervals between consecutive pairs of onsets,
measuring clockwise arc-lengths or distances around the circle
of circumference~$n$.
The third representation of rhythms and scales writes
the onsets as a \emph{subset} of the set of all pulses,
numbered $0, 1, \dots, n-1$, with a subscript of $n$ on the right-hand side
of the subset to denote the timespan.
Clough and Douthett~\cite{clough-91} use this notation to represent scales.
For example, the Cuban clave Son rhythm can be represented as
\rbox{[x . . x . . x . . . x . x . . .]} in box-like notation,
$(3,3,4,2,4)$ in clockwise distance sequence notation, and
$\{0,3,6,10,12\}_{16}$ in subset notation.
Finally, the fourth representation is a graphical
\emph{clock diagram}~\cite{toussaint-02}, such as Figure~\ref{even-cuban},
in which the zero label denotes the start of the rhythm
and time flows in a clockwise direction.
In such clock diagrams we usually connect adjacent onsets by
line segments, forming a polygon.

\paragraph{Even Rhythms.}

Consider the following three 12/8-time rhythms expressed
in box-like notation: 
\rbox{[x . x . x . x . x . x .]}, 
\rbox{[x . x . x x . x . x . x]}, and \rbox{[x . . . x x . . x x x .]}.
It is intuitively clear that the first rhythm is more \emph{even} (well spaced)
than the second rhythm, and that the second rhythm is more even than the
third rhythm.
In fact, the second rhythm is the internationally most
well known of all African timelines.  It is traditionally
played on an iron bell, and is known on the world scene mainly
by its Cuban name \emph{Bemb\'e}~\cite{toussaint-03}.
Traditional rhythms tend to exhibit such properties
of evenness to some degree. 

Why do many traditional rhythms display such evenness?
Many are timelines (also sometimes called \emph{claves}), that is,
rhythms repeated throughout a piece that
serve as a rhythmic reference point~\cite{uribe-96, ortiz-95}. 
Often these claves have a \emph{call-and-response structure}, meaning
that the pattern is divided into two parts:
the first poses a rhythmic question, usually by creating rhythmic tension,
and the second part answers this question by releasing that tension. 
A good example of this structure is the popular clave Son
\rbox{[x . . x . . x . . . x . x . . .]}.
This clave creates such tension through syncopation,
which can be found between the second and third onsets
as well as between the third and fourth onsets.
The latter is weak syncopation because the strong beat at position $8$
lies half-way between the third and fourth onsets.
(The strong beats of the underlying 4/4 meter (beat) 
occur at positions $0$, $4$, $8$, and~$12$.)
On the other hand, the former syncopation is strong
because the strong beat at position $4$ is closer to the second onset than
to the third onset~\cite{gomez-05}.
Claves played with instruments that produce unsustained notes 
often use syncopation and accentuation to bring about rhythmic tension. 
Many clave rhythms create syncopation by evenly distributing 
onsets in contradiction with the pulses of the underlying meter.
For example, in the clave Son, the first three onsets are equally spaced
at the distance of three sixteenth pulses, which forms a contradiction
because $3$ does not divide~$16$.
Then, the response of the clave answers with an offbeat onset,
followed by an onset on the fourth strong beat of a 4/4 meter,
releasing that rhythmic tension.

On the other hand, a rhythm should not be too even, such as the example
\rbox{[x . x . x . x . x . x .]}.
Indeed, in the most interesting rhythms with $k$ onsets and timespan~$n$,
$k$~and $n$ are relatively prime (have no common divisor larger than~$1$).
This property is natural because the rhythmic contradiction is easier
to obtain if the onsets do not coincide with the strong beats of the meter. 
Also, we find that many claves have an onset on the last
strong beat of the meter, as does the clave Son.
This is a natural way to respond in the call-and-response structure. 
A different case is that of the Bossa-Nova clave 
\rbox{[x . . x . . x . . . x . . x . .]}. 
This clave tries to break the feeling of the pulse and, 
although it is very even, it produces a cycle that perceptually does not coincide 
with the beginning of the meter.  

This prevalence of evenness in world rhythms has led to the study of
mathematical measures of evenness
in the new field of
mathematical ethnomusicology~\cite{chemillier-02, toussaint-04b, toussaint-05},
where they may help to identify, if not explain, cultural
preferences of rhythms in traditional music.
Furthermore, evenness in musical chords
plays a significant role in the efficacy of voice leading as discussed in
the work of Tymoczko~\cite{tymoczko}.

The notion of \emph{maximally even sets} with respect to scales represented on
a circle was introduced by Clough and Douthett~\cite{clough-91}.
According to Block and Douthett~\cite{block-94},
Douthett and Entringer went further by constructing several mathematical
measures of the amount of \emph{evenness} contained in a scale;
see \cite[page~40]{block-94}.
One of their evenness measures simply sums the interval arc-lengths
(geodesics along the circle) between all pairs of onsets
(or more precisely, onset points).
This measure differentiates between rhythms
that differ widely from each other.
For example, the two four-onset rhythms
\rbox{[x . . . x . . .  x . . .  x . . .]}
and \rbox{[x . x . x . . x . . . . . . . .]} yield evenness values of 32
and 23, respectively, reflecting clearly that the first rhythm is more
evenly spaced than the second.
However, the measure is too coarse to be useful
for comparing rhythm timelines such as those studied
in~\cite{toussaint-02, toussaint-03}.
For example, all six fundamental 4/4-time clave/bell patterns discussed in
\cite{toussaint-02} and shown in Figure~\ref{even-cuban} have an equal
pairwise sum of geodesic distances, namely 48, yet the \emph{Bossa-Nova} clave
is intuitively more even than, say, the \emph{Soukous} and \emph{Rumba} claves.

Another distance measure that has been considered is the sum of pairwise
chordal distances between adjacent onsets,
measured by Euclidean distance between points on the circle.
It can be shown that the rhythms maximizing this measure of evenness
are precisely the rhythms with maximum possible area.
Rappaport \cite{rappaport-05} shows that many of the most common chords and
scales in Western harmony correspond to these maximum-area sets.
This evenness measure is finer than the sum of pairwise arc-lengths,
but it still does not distinguish half the rhythms in
Figure~\ref{even-cuban}.  Specifically, the Son, Rumba, and
Gahu claves have the same occurrences of arc-lengths between consecutive onsets,
so they also have the same occurrences (and hence total) of distances  
between consecutive onsets.

%

The measure of evenness we consider here is the sum of all pairwise
Euclidean distances between
points on the circle, as described by Block and Douthett~\cite{block-94}.
This measure is more discriminating than the others,
and is therefore the preferred measure of evenness. 
For example, this measure distinguishes all of the six rhythms
in Figure~\ref{even-cuban}, ranking the Bossa-Nova rhythm as the most even,
followed by the Son, Rumba, Shiko, Gahu, and Soukous.
Intuitively, the rhythms with a larger sum of pairwise chordal distances
have more ``well spaced'' onsets.

%
%
%
%
%
%

In Section~\ref{sec:Even}, we study the mathematical and computational aspects
of rhythms that maximize evenness.  We describe three algorithms that generate
such rhythms, show that these algorithms are equivalent, and show that in fact
the rhythm of maximum evenness is essentially unique.
These results characterize rhythms with maximum evenness.
One of the algorithms is the Euclidean-like algorithm from
Section~\ref{sec:Euclid}, proving that the rhythms of maximum evenness
are precisely the Euclidean rhythms from that section.

\paragraph{Deep Rhythms.}
Another important property of rhythms and scales that we study in this paper
is \emph{deepness}.
Consider a rhythm with $k$ onsets and timespan~$n$,
represented as a set of $k$ points on a circle of circumference~$n$.
Now measure the arc-length/geodesic distances along the circle
between all pairs of onsets.
%
%
A musical scale or rhythm is \emph{Winograd-deep}
if every distance $1, 2, \dots, \lfloor n/2 \rfloor$
has a unique number of occurrences
(called the \emph{multiplicity} of the distance).
For example, the rhythm \rbox{[x x x . x .]} is Winograd-deep because distance
$1$ appears twice, distance $2$ appears thrice, and distance $3$ appears once.

The notion of deepness in scales was introduced by Winograd in an oft-cited
but unpublished class project report from 1966 \cite{Winograd-1966},
disseminated and further developed by the class instructor Gamer in 1967
\cite{Gamer-1967b, Gamer-1967}, and considered further in numerous papers
and books, e.g., \cite{clough-99b, Johnson-2003}.
Equivalently, a scale is \emph{Winograd-deep} if the number of onsets it has
in common with each of its cyclic shifts (rotations) is unique.
This equivalence is the Common Tone Theorem \cite[page~42]{Johnson-2003},
and it is originally described by Winograd~\cite{Winograd-1966}
(who in fact uses this definition as his primary definition of ``deep'').
Deepness is one property of the ubiquitous Western \emph{diatonic} 12-tone
major scale \rbox{[x . x . x x . x . x . x]} \cite{Johnson-2003},
and it captures some of the rich
structure that perhaps makes this scale so attractive.

\begin{figure}
  \centering
  \includegraphics[scale=0.6]{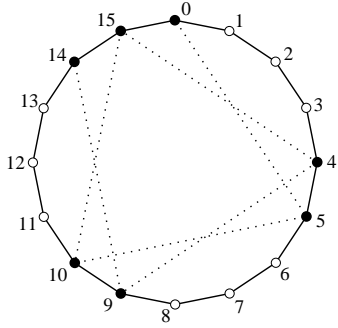}
  \caption{A rhythm with $k=7$ onsets and timespan $n=16$ that is
    Winograd-deep and thus Erd\H{o}s-deep. 
    Distances ordered by multiplicity from $1$ to $6$ are
    $2$, $7$, $4$, $1$, $6$, and~$5$.
    The dotted line shows how the rhythm is generated by multiples of $m=5$.}
  \label{deep example}
\end{figure}

Winograd-deepness translates directly from scales to rhythms.
For example, the diatonic major scale is equivalent to the famous Cuban rhythm 
\emph{Bemb\'e} \cite{pressing-83,toussaint-03}.
Figure~\ref{deep example} shows a graphical example of a Winograd-deep rhythm.
However, the notion of Winograd-deepness is rather restrictive for rhythms,
because it requires half of the pulses in a timespan
(rounded to a nearest integer) to be onsets.
In contrast, for example, the popular Bossa-Nova rhythm
\rbox{[x . . x . . x . . . x . . x . .]} $ = \{0,3,6,10,13\}_{16}$ 
pictures in Figure~\ref{even-cuban} 
has only five onsets in a timespan of sixteen. 
Nonetheless, if we focus on just the distances that appear at least once
between two onsets, then the multiplicities of occurrence are all unique and
form an interval starting at~$1$: distance $4$ occurs once,
distance $7$ occurs twice, distance $6$ occurs thrice, and
distance $3$ occurs four times.

We therefore define a rhythm (or scale) to be \emph{Erd\H{o}s-deep}
if it has $k$ onsets and, for every multiplicity $1, 2, \dots, k-1$,
there is a nonzero arc-length/geodesic distance determined by the points
on the circle with exactly that multiplicity.
The same definition is made by Toussaint~\cite{Toussaint-2004-CGW}.
Every Winograd-deep rhythm is also Erd\H{o}s-deep, so this definition
is strictly more general.

To further clarify the difference between Winograd-deep and Erd\H{o}s-deep
rhythms, it is useful to consider which distances can appear.
For a rhythm to be Winograd-deep, all the distances between $1$ and $k-1$
must appear a unique number of times.  In contrast, to be an Erd\H{o}s-deep
rhythm, it is only required that any distance that appears must have a unique
multiplicity.
Thus, the Bossa-Nova rhythm is not Winograd-deep because distances
$1, 2$ and $5$ do not appear.




The property of Erd\H{o}s deepness involves only the distances between points
in a set, and is thus a feature of \emph{distance geometry}---in this case,
in the discrete space of $n$ points equally spaced around a circle.
In 1989, Paul Erd\H{o}s~\cite{Erdos-1989} considered the analogous question
in the plane, asking whether there exist $n$ points in the plane
(no three on a line and no four on a circle) such that,
for every $i = 1, 2, \dots, n-1$, there is a distance determined
by these points that occurs exactly $i$~times.
Solutions have been found for $n$ between $2$ and~$8$,
but in general the problem remains open.
Pal\'asti~\cite{Palasti-1989} considered a variant of this problem
with further restrictions---no three points form a regular triangle,
and no one is equidistant from three others---and solved it for $n = 6$.

In Section~\ref{sec:Deep}, we characterize all rhythms that are Erd\H{o}s-deep.
In particular, we prove that all deep rhythms, besides one exception,
are \emph{generated}, meaning that the rhythm can be represented as
$\{0, m, 2 m, \dots, (k-1) m\}_n$ for some integer~$m$,
where all arithmetic is modulo~$n$.
In the context of scales, the concept of ``generated'' was defined
by Wooldridge~\cite{wooldridge-1993} 
and used by Clough et al.~\cite{clough-99b}.
For example, the rhythm in Figure~\ref{deep example} is generated with $m = 5$. 
Our characterization generalizes a similar characterization for
Winograd-deep scales proved by Winograd \cite{Winograd-1966},
and independently by Clough et al.~\cite{clough-99b}.

In the pitch domain, generated scales are very common. 
The Pythagorean tuning is a good example:
all its pitches are generated from the fifth of ratio $3:2$ modulo the octave. 
Another example is the equal-tempered scale, 
which is generated with a half-tone of ratio $\sqrt[12]{2}$ \cite{barbour-04}. 
Generated scales are also of interest in the theory of the
well-formed scales~\cite{carey-98}.

Generated rhythms have an interesting property called \emph{shellability}.
If we remove the ``last'' generated onset $14$ from the rhythm in 
Figure~\ref{deep example}, the resulting rhythm is still generated,
and this process can be repeated until we run out of onsets.
In general, every generated rhythm has a \emph{shelling} in the sense that
it is always possible to remove a particular onset and obtain another
generated rhythm.

Shellings of rhythms play an important role in musical improvisation.
For example, most African drumming music consists of rhythms operating
on three different strata: the unvarying timeline usually
provided by one or more bells, one or more rhythmic motifs
played on drums, and an improvised solo (played by the lead drummer)
riding on the other rhythmic structures.
Shellings of rhythms are relevant to the improvisation of
solo drumming in the context of such a rhythmic background.
The solo improvisation must respect the style and feeling
of the piece which is usually determined by the timeline.
A common technique to achieve this effect is to ``borrow'' notes
from the timeline, and to alternate
between playing subsets of notes from the timeline
and from other rhythms that interlock with the 
timeline~\cite{anku-97, agawu-86}.
In the words of Kofi Agawu~\cite{agawu-86}, 
``It takes a fair amount of expertise
to create an effective improvisation that is at the same time
stylistically coherent''.
The borrowing of notes from the timeline may be regarded as
a fulfillment of the requirements of style coherence.

Of course, some subsets of notes of a rhythm may be better
choices than others.
For example, it seems reasonable that, if a rhythm is deep,
one should select subsets of the rhythm that are also deep.
Furthermore, a shelling seems a natural way
to decrease or increase the density of the notes
in an improvisation that respects these constraints.
For example, in the \emph{Bemb{\'e}} bell timeline
\rbox{[x.x.xx.x.x.x]}, which is deep,
one possible shelling is
\rbox{[x.x.xx.x.x..]},
\rbox{[x.x.x..x.x..]},
\rbox{[x.x....x.x..]},
\rbox{[x.x....x....]}.
All five rhythms sound good and are stylistically coherent.
To our knowledge, shellings have not been studied from the
musicological point of view.
However, they may be useful both for theoretical analysis
as well as providing formal rules for ``improvisation'' techniques.

One of the consequences of our characterization that we obtain
in Section~\ref{sec:Deep} is that every Erd\H{o}s-deep rhythm
has a shelling.  More precisely, it is always possible to remove a particular
onset that preserves the Erd\H{o}s-deepness property.
Finally, to tie everything together,
we show that essentially all Euclidean rhythms (or equivalently,
rhythms that maximize evenness) are Erd\H{o}s-deep.


\section{Euclid and Evenness in Various Disciplines}
\label{sec:Euclid}

In this section, we first describe Euclid's classic algorithm
for computing the greatest common divisor of two integers.
Then, through an unexpected connection to timing systems
in neutron accelerators, we see how the same type of algorithm
can be used as an approach to maximizing ``evenness'' in a binary string
with a specified number of zeroes and ones.
This algorithm defines an important family of rhythms,
called \emph{Euclidean rhythms}, which we show appear throughout world music.
Finally, we see how similar ideas have been used in algorithms
for drawing digital straight lines and in combinatorial strings
called Euclidean strings.


\subsection{The Euclidean Algorithm for Greatest Common Divisors}
\label{Euclid gcd}

The Euclidean algorithm for computing the 
greatest common divisor of two integers is one of 
the oldest known algorithms (circa 300~B.C.). It 
was first described by Euclid in Proposition~2 of 
Book~VII of \emph{Elements}~\cite{euclid-56, franklin-56}.
%
%
Indeed, Donald Knuth~\cite{knuth-98} calls this algorithm the
``granddaddy of all algorithms, because it is the oldest nontrivial
algorithm that has survived to the present day''.

The idea of the algorithm is simple:
repeatedly replace the larger of the two numbers by their difference
until both are equal.
This final number is then the greatest common divisor.
For example, consider the numbers $5$ and~$13$.
First, $13-5 = 8$; then $8-5=3$; next $5-3=2$; then $3-2=1$;
and finally $2-1=1$. 
Therefore, the greatest common divisor of $5$ and $13$ is~$1$;
in other words, $5$ and $13$ are relatively prime.

The algorithm can also be described succinctly in a recursive
manner as follows~\cite{cormen-01}. 
Let $k$ and $n$ be the input integers with $k < n$.
\begin{algorithmbox}{\textsc{Euclid}$(k,n)$}
\begin{enumerate*}
\item {\bf if} $k=0$ {\bf then return} $n$
\item {\bf else return} \textsc{Euclid}$(n \bmod k, k)$
\end{enumerate*}
\end{algorithmbox}


Running this algorithm with $k=5$ and $n=13$, we obtain
\textsc{Euclid}$(5,13) = $ \textsc{Euclid}$(3,5) = $
\textsc{Euclid}$(2,3) = $ \textsc{Euclid}$(1,2) = $
\textsc{Euclid}$(0,1) = 1$.
Note that this division version of Euclid's algorithm
skips one of the steps $(5,8)$ made by the original subtraction version.


\subsection{Evenness and Timing Systems in Neutron Accelerators}
\label{Bjorklund algorithm}


One of our main musical motivatations is to find rhythms with a specified
timespan and number of onsets that maximize evenness.
Bjorklund~\cite{bjorklund-03a, bjorklund-03b} was faced with a similar problem
of maximizing evenness, but in a different context:
the operation of components such as high-voltage power supplies
of spallation neutron source (SNS) accelerators used in nuclear physics.
In this setting, a timing system controls a collection of gates
over a time window divided into $n$ equal-length intervals.
(In the case of SNS, each interval is 10 seconds.)
The timing system can send signals to enable a gate
during any desired subset of the $n$ intervals.
For a given number $n$ of time intervals, and
another given number $k < n$ of signals,
the problem is to distribute the pulses
as evenly as possible among these $n$ intervals.
Bjorklund~\cite{bjorklund-03a} represents this problem as
a binary sequence of $k$ ones and $n-k$ zeroes, where each
bit represents a time interval and the ones represent
the times at which the timing system sends a signal.
The problem then reduces to the following:
construct a binary sequence of $n$ bits with $k$ ones
such that the $k$ ones are distributed as evenly as possible
among the $(n-k)$ zeroes.

One simple case is when $k$ evenly divides~$n$ (without remainder),
in which case we should place ones every $n/k$ bits.
For example, if $n=16$ and $k=4$, then the solution is
\mbox{[1000100010001000]}.
This case corresponds to $n$ and $k$ having a common divisor of~$k$.
More generally, if the greatest common divisor between $n$ and $k$ is~$g$,
then we would expect the solution to decompose into $g$ repetitions of a
sequence of $n/g$ bits.  Intuitively, a string of maximum evenness
should have this kind of symmetry, in which it decomposes into
more than one repetition, whenever such symmetry is possible.
This connection to greatest common divisors suggests that a rhythm
of maximum evenness might be computed using an algorithm like Euclid's.
Indeed, Bjorklund's algorithm closely mimics the structure Euclid's algorithm,
although this connection has never been mentioned before.


We describe Bjorklund's algorithm by using one of his examples.
Consider a sequence with $n=13$ and $k=5$.
Because $13-5=8$, we start by considering a sequence consisting of 5 ones
followed by 8 zeroes which should be thought of as 13 sequences of
one bit each:
\begin{center}
\mbox{[1][1][1][1][1][0][0][0][0][0][0][0][0]}
\end{center}
If there is more than one zero the algorithm moves zeroes in stages.
We begin by taking zeroes one at a time (from right to left),
placing a zero after each one (from left to right), to produce five
sequences of two bits each, with three zeroes remaining:
\begin{center}
\mbox{[10] [10] [10] [10] [10]} \mbox{[0] [0] [0]}
\end{center}
Next we distribute the three remaining zeros in a similar manner,
by placing a \mbox{[0]} sequence after each \mbox{[10]} sequence:
\begin{center}
\mbox{[100] [100] [100] [10] [10]}
\end{center}
Now we have three sequences of three bits each, and a remainder of two
sequences of two bits each. Therefore we continue in the same manner,
by placing a \mbox{[10]} sequence after each \mbox{[100]} sequence:
\begin{center}
\mbox{[10010] [10010] [100]}
\end{center}
The process stops when the remainder consists of only one sequence
(in this case the sequence \mbox{[100]}), or we run out of zeroes
(there is no remainder).
The final sequence is thus the concatenation of \mbox{[10010]},
\mbox{[10010]}, and \mbox{[100]}:
\begin{center}
\mbox{[1001010010100]}
\end{center}

We could proceed further in this process by inserting
\mbox{[100]} into \mbox{[10010] [10010]}. However, Bjorklund argues that,
because the sequence is cyclic, it does not matter (hence his stopping rule).
For the same reason, if the initial sequence has a group of ones
followed by only one zero,
the zero is considered as a remainder consisting of one
sequence of one bit, and hence nothing is done.
Bjorklund~\cite{bjorklund-03a} shows that the final sequence may be computed
from the initial sequence using $O(n)$ arithmetic operations in the worst case.

A more convenient and visually appealing way to implement this algorithm by
hand is to perform the sequence of insertions in a vertical manner as follows.
First take five zeroes from the right and place them under the five ones on the
left:
\begin{center}
\begin{tabular}{l}
  \mbox{1 1 1 1 1 0 0 0} \\
  \mbox{0 0 0 0 0}
\end{tabular}
\end{center}
Then move the three remaining zeroes in a similar manner:
\begin{center}
\begin{tabular}{l}
1 1 1 1 1 \\
0 0 0 0 0 \\
0 0 0
\end{tabular}
\end{center}
Next place the two remainder columns on the right
under the two leftmost columns:
\begin{center}
\begin{tabular}{l}
1 1 1 \\
0 0 0 \\
0 0 0 \\
1 1 \\
0 0
\end{tabular}
\end{center}
Here the process stops because the remainder consists of only one column.
The final sequence is obtained by concatenating the three columns
from left to right:
\begin{center}
1 0 0 1 0 1 0 0 1 0 1 0 0
\end{center}

Bjorklund's algorithm applied to a string of $n$ bits
consisting of $k$ ones and $n-k$ zeros has the same structure as
running \textsc{Euclid}$(k,n)$.
Indeed, Bjorklund's algorithm uses the repeated subtraction form of
division, just as Euclid did in his \emph{Elements}~\cite{euclid-56}.
It is also well known that applying the algorithm \textsc{Euclid}$(k,n)$
to two $O(n)$ bit numbers (binary sequences of length~$n$) causes it to
perform $O(n)$ arithmetic operations in the worst case~\cite{cormen-01}.

\subsection{Euclidean Rhythms}
\label{Euclidean rhythms}

The binary sequences generated by Bjorklund's algorithm,
as described in the preceding, may be considered as one family
of rhythms.
Furthermore, because Bjorklund's algorithm is a way of visualizing
the repeated-subtraction version of the Euclidean algorithm, we call
these rhythms \emph{Euclidean rhythms}.
We denote the Euclidean rhythm by $E(k,n)$,
where $k$ is the number of ones (onsets) and
$n$ (the number of pulses) is the length of the sequence (zeroes plus ones).
For example, $E(5,13) = $ [1001010010100].
The zero-one notation is not ideal for representing binary rhythms
because it is difficult to visualize the locations of the onsets
as well as the duration of the inter-onset intervals.
In the more iconic box notation, the preceding rhythm is written as
\mbox{$E(5,13) = $ \rbox{[x . . x . x . . x . x . .]}}.

\begin{figure}
  \centering
  \includegraphics*[scale=0.6]{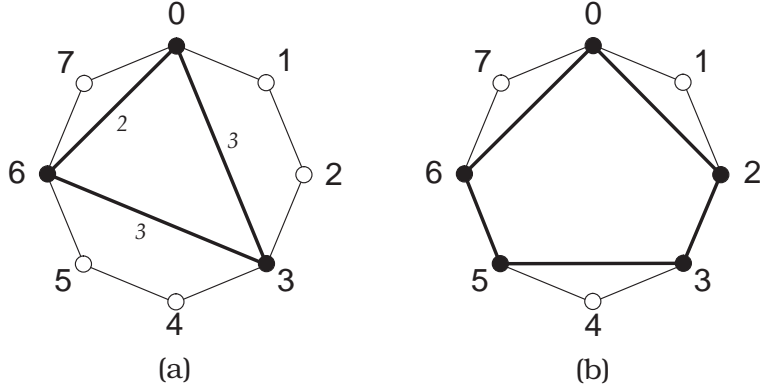}
  \caption{{(a) The Euclidean rhythm $E(3,8)$ is the Cuban \emph{tresillo}.
            (b) The Euclidean rhythm $E(5,8)$ is the Cuban \emph{cinquillo}.}}
  \label{tresillo-cinquillo-polygons}
\end{figure}

The rhythm $E(5,13)$ is in fact used in Macedonian music~\cite{arom-04},
but having a timespan of~$13$ (and defining a measure of length~$13$),
it is rarely found in world music.
For contrast, let us consider two widely used values of $k$ and~$n$;
in particular, what is $E(3,8)$?
Applying Bjorklund's algorithm to the corresponding sequence
\mbox{[11100000]}, the reader may easily verify that the resulting
Euclidean rhythm is $E(3,8) = $ \rbox{[x . . x . . x .]}.
Figure~\ref{tresillo-cinquillo-polygons}(a) shows a clock diagram of
this rhythm, where the numbers by the sides of the triangle
indicate the arc-lengths between those onsets.

The Euclidean rhythm $E(3,8)$ is one of the most famous on the planet.
In Cuba, it goes by the name of the \emph{tresillo}, and in the USA,
it is often called the \emph{Habanera} rhythm.
It was used in hundreds of \emph{rockabilly} songs during the 1950's.
It can often be heard in early rock-and-roll hits in the left-hand patterns
of the piano, or played on the string bass or 
saxophone~\cite{brewer-99, floyd-99, morrison-96}.
A good example is the bass rhythm in Elvis Presley's
\emph{Hound Dog}~\cite{brewer-99}.
The tresillo pattern is also found widely in West African traditional music.
For example, it is played on the \emph{atoke} bell in the \emph{Sohu},
an \emph{Ewe} dance from Ghana~\cite{kauffman-80}.
The tresillo can also be recognized as the first bar (first eight pulses)
of the ubiquitous two-bar clave Son shown in Figure~\ref{even-cuban}(b).

In the two examples $E(5,13)$ and $E(3,8)$, there are fewer ones than zeros.
If instead there are more ones than zeros, Bjorklund's algorithm yields the
following steps with, for example, $k=5$ and $n=8$:

\begin{center}
\mbox{[1 1 1 1 1 0 0 0]}

\mbox{[10] [10] [10] [1] [1]}

\mbox{[101] [101] [10]}

\mbox{[1 0 1 1 0 1 1 0]}
\end{center}

The resulting Euclidean rhythm is $E(5,8) = $ \rbox{[x . x x . x x .]}.
Figure~\ref{tresillo-cinquillo-polygons}(b) shows a clock diagram for
this rhythm.
It is another famous rhythm on the world scene.
In Cuba, it goes by the name of the \emph{cinquillo}
and it is intimately related to the tresillo~\cite{floyd-99}.
It has been used in jazz throughout the 20th century~\cite{rahn-96},
and in rockabilly music.
For example, it is the hand-clapping pattern in
Elvis Presley's \emph{Hound Dog}~\cite{brewer-99}.
The cinquillo pattern is also widely used in West African
traditional music~\cite{rahn-87,toussaint-02}, as well as
Egyptian~\cite{hagoel-03} and Korean~\cite{hye-ku-81} music.


We show in this paper that Euclidean rhythms have two important
properties: they maximize evenness and they are deep.
The evenness property should come as no surprise,
given how we designed the family of rhythms.
To give some feeling for the deepness property, we consider the
two examples in Figure~\ref{tresillo-cinquillo-polygons},
which have been labeled with the distances between all pairs of onsets,
measured as arc-lengths.
The \emph{tresillo} in Figure~\ref{tresillo-cinquillo-polygons}(a)
has one occurrence of distance~$2$ and two occurrences of distance~$3$.
The \emph{cinquillo} in Figure~\ref{tresillo-cinquillo-polygons}(b)
contains one occurrence of distance~$4$, two occurrences of distance~$1$,
three occurrences of distance~$2$, and four occurrences of distance~$3$.
Thus, every distance has a unique multiplicity,
making these rhythms Erd\H{o}s-deep.


\subsection{Euclidean Rhythms in Traditional World Music}

\begin{figure}
  \centering
  \includegraphics*[scale=0.6]{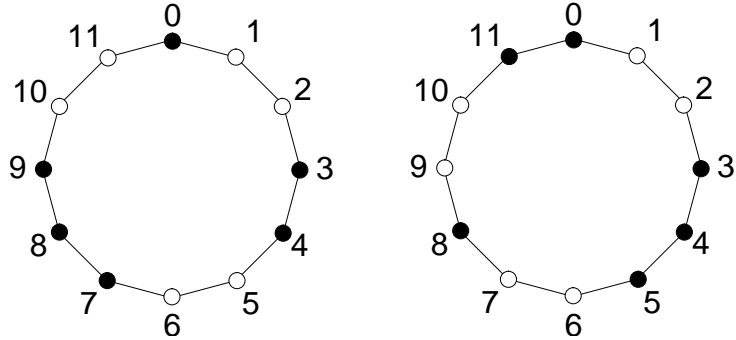}
  \caption{{These two rhythms are instances of the same rhythm necklace.}}
  \label{necklaces}
\end{figure}

In this section, we list all the Euclidean rhythms found in world music 
that we have collected so far, restricting attention to those in which 
$k$ and $n$ are relatively prime.
In some cases, the Euclidean rhythm is a rotated version of a commonly
used rhythm.  If a rhythm is a rotated version of another, we say
that they are instances of the same \emph{necklace}.
Thus a rhythm necklace is a clockwise distance sequence that disregards
the starting point in the cycle.
Figure~\ref{necklaces} illustrates an example of two rhythms
that are instances of the same necklace.

Rhythms in which $k$ and $n$ have a common divisor larger than~$1$
are common all over the planet in traditional, classical, and
popular genres of music.
For example, $E(4,12) = $ \rbox{[x . . x . . x . . x . .]}
is the 12/8-time \emph{Fandango} clapping pattern in the Flamenco music of
southern Spain, where `\rbox{x}' denotes a loud clap and `\rbox{.}' denotes
a soft clap~\cite{banez-04}.
However, the string itself is periodic: $E(4,12)$ has period~$3$,
even though it appears in a timespan of~$12$.
For this reason, we restrict ourselves to the more interesting
Euclidean rhythms that do not decompose into repetitions of
shorter Euclidean rhythms.
We are also not concerned with rhythms that have only one onset
(\rbox{[x .]}, \rbox{[x . .]}, etc.), and similarly with any repetitions
of these rhythms (for example, \rbox{[x . x .]}).


There are surprisingly many Euclidean rhythms with $k$ and $n$ relatively prime
that are found in world music.
The following list includes more than 40 such rhythms uncovered so far.

\medskip

{$E(2,3) = $ \rbox{[x x .]}} $ = (12)$
is a common Afro-Cuban drum pattern
when started on the second onset as in \rbox{[x . x]}.
For example, it is the conga rhythm of the (6/8)-time
\emph{Swing Tumbao}~\cite{klower-97}.
It is common in Latin American music, as for example in the
\emph{Cueca}~\cite{vanderlee-95}, and the
\emph{coros de clave}~\cite{rodriguez-97}.
It is common in Arabic music, as for example in the \emph{Al T\'aer} rhythm
of Nubia~\cite{hagoel-03}.
It is also a rhythmic pattern of the Drum Dance of the Slavey Indians
of Northern Canada~\cite{asch-75}.

\mbox{$E(2,5) = $ \rbox{[x . x . .]}} $ = (23)$
is a rhythm found in Greece, Namibia, Rwanda
and Central Africa~\cite{arom-04}.
It is also a 13th century Persian rhythm
called \emph{Khafif-e-ramal}~\cite{wright-78}, as well
as the rhythm of the Macedonian dance \emph{Makedonka}~\cite{singer-74}.
Tchaikovsky used it as the metric pattern in the second movement
of his \emph{Symphony No.\ 6}~\cite{keith-91}.
Started on the second onset as in \rbox{[x . . x .]}
it is a rhythm found in Central Africa, Bulgaria, Turkey,
Turkestan and Norway~\cite{arom-04}.
It is also the metric pattern of Dave Brubeck's \emph{Take Five}, as well
as \emph{Mars} from \emph{The Planets} by Gustav Holst~\cite{keith-91}.
Both starting points determine metric patterns used in Korean
music~\cite{hye-ku-81}.

\mbox{$E(3,4) = $ \rbox{[x x x .]}} $ = (112)$
is a pattern used in the \emph{Baia\'o} rhythm
of Brazil~\cite{uribe-93}, as well as the \emph{polos} of Bali~\cite{montfort-85}.
Started on the second onset, as in \rbox{[x x . x]}, it is the
\emph{Catarete} rhythm of the indigenous people of Brazil~\cite{uribe-93}.
Started on the third onset, as in \rbox{[x . x x]}, it is the
archetypal pattern of the \emph{Cumbia} from
Colombia~\cite{manuel-85}, as well as a \emph{Calypso}
rhythm from Trinidad~\cite{evans-66}.
It is also a 13th century Persian rhythm called
\emph{Khalif-e-saghil}~\cite{wright-78}, as well as the \emph{trochoid choreic}
rhythmic pattern of ancient Greece~\cite{mathiesen-85}.
Started on the silent note (anacrusis), as in
\rbox{[. x x x]}, it is a popular flamenco rhythm
used in the \emph{Taranto}, the \emph{Tiento}, the \emph{Tango},
and the \emph{Tanguillo}~\cite{gamboa-02}.
It is also the \emph{Rumba} clapping pattern in flamenco,
as well as a second pattern used in the \emph{Baia\'o} rhythm
of Brazil~\cite{uribe-93}.

\mbox{$E(3,5) = $ \rbox{[x . x . x]}} $ = (221)$,
when started on the second onset,
is another 13th century Persian rhythm
by the name of \emph{Khafif-e-ramal}~\cite{wright-78},
as well as a Romanian folk-dance rhythm~\cite{proca-69},
and the \emph{Sangsa Py\v{o}lgok} drum pattern in Korean
music~\cite{hye-ku-81}.

\mbox{$E(3,7) = $ \rbox{[x . x . x . .]}} $ = (223)$
is a rhythm found in Greece, Turkestan,
Bulgaria, and Northern Sudan~\cite{arom-04}.
It is the \emph{D\'awer turan} rhythmic pattern of Turkey~\cite{hagoel-03}.
It is the \emph{Ruchenitza} rhythm
used in a Bulgarian folk-dance~\cite{pressing-83}, as well as
the rhythm of the Macedonian dance \emph{Eleno Mome}~\cite{singer-74}.
It is also the rhythmic pattern of
Dave Brubeck's \emph{Unsquare Dance}, and
Pink Floyd's \emph{Money}~\cite{keith-91}.
Started on the second onset, as in \rbox{[x . x . . x .]},
it is a Serbian rhythm~\cite{arom-04}.
Started on the third onset, as in \rbox{[x . . x . x .]},
it is a rhythmic pattern found in Greece and Turkey~\cite{arom-04}.
In Yemen it goes by the name of \emph{Daasa al zreir}~\cite{hagoel-03}.
It is also the rhythm of the
Macedonian dance \emph{Tropnalo Oro}~\cite{singer-74},
the rhythm for the Bulgarian \emph{Makedonsko Horo} dance~\cite{wade-04},
as well as the meter and clapping pattern of the \emph{t\={\i}vr\={a} t\={a}l}
of North Indian music~\cite{clayton-00}.

\mbox{$E(3,8) = $ \rbox{[x . . x . . x .]}} $ = (332)$
is the Cuban \emph{tresillo} pattern discussed
in the preceding~\cite{floyd-99}, the most important traditional bluegrass banjo
rhythm~\cite{keith-91},
as well as the \emph{Mai} metal-blade
pattern of the \emph{Aka} Pygmies~\cite{arom-91}.
It is common in West Africa~\cite{brandel-59} and many other parts of the world
such as Greece and Northern Sudan~\cite{arom-04}.
Curt Sachs~\cite{sachs-53} and Willi Apel~\cite{apel-60}
consider it to be one of the most important rhythms in Renaissance music.
Indeed, it dates back to the Ancient Greeks who called it the
\emph{dochmiac} pattern~\cite{brandel-59}.
In India it is one of the 35 s\={u}l\={a}di t\={a}las of Karnatak
music~\cite{london-04}.
Started on the second onset, it is a drum pattern of the
\emph{Samhy\v{o}n Tod\v{u}ri} Korean instrumental music~\cite{hye-ku-81}.
It is also found in Bulgaria and Turkey~\cite{arom-91}.
Started on the third onset, it is the \emph{Nandon Bawaa} bell
pattern of the \emph{Dagarti} people of northwest Ghana~\cite{hartigan-95},
and is also found in Namibia and Bulgaria~\cite{arom-91}.

\mbox{$E(3,11) = $ \rbox{[x . . . x . . . x . .]} $ = (443)$}
is the metric pattern
of the \emph{sav\={a}r\={\i} t\={a}l}
of North Indian music~\cite{clayton-00}.

\mbox{$E(3,14) = $ \rbox{[x . . . . x . . . . x . . .]} $ = (554)$}
is the clapping pattern
of the \emph{dham\={a}r t\={a}l}
of North Indian music~\cite{clayton-00}.

\mbox{$E(4,5) = $ \rbox{[x x x x .]} $ = (1112)$}
is the rhythmic pattern of the \emph{Mirena}
rhythm of Greece~\cite{hagoel-03}. Started on the fourth onset,
as in \rbox{[x . x x x]}, it is the \emph{Tik} rhythm of Greece~\cite{hagoel-03}.

\mbox{$E(4,7) = $ \rbox{[x . x . x . x]} $ = (2221)$}
is another \emph{Ruchenitza}
Bulgarian folk-dance rhythm~\cite{pressing-83}.
Started on the third onset, it is the \emph{Kalam\'atianos}
Greek dance rhythm~\cite{hagoel-03}, as well as the \emph{Shaigie}
rhythmic pattern of Nubia~\cite{hagoel-03}.
Started on the fourth (last) onset, it is the rhythmic
pattern of the \emph{Dar daasa al mutawasit} of Yemen~\cite{hagoel-03}.

\mbox{$E(4,9) = $ \rbox{[x . x . x . x . .]} $ = (2223)$}
is the \emph{Aksak} rhythm
of Turkey~\cite{brailoiu-51} (also found in Greece) as well as the rhythm of the
Macedonian dance \emph{Kambani Bijat Oro}~\cite{singer-74} and
the Bulgarian dance \emph{Daichovo Horo}~\cite{rice-04}.
In Bulgarian music fast tunes with this metric pattern are called
\emph{Dajchovata} whereas slow tunes with this same pattern
are called \emph{Samokovskata}~\cite{rice-80}.
It is the rhythmic ostinato of a lullaby discovered by Simha Arom
in south-western Za\"{\i}re~\cite{arom-04}.
It is the metric pattern used by Dave Brubeck in his well known piece
\emph{Rondo a la Turk}~\cite{keith-91}.
When it is started on the second onset, as in \rbox{[x . x . x . . x .]},
it is found in Bulgaria and Serbia~\cite{arom-04}.
Started on the third onset, as in \rbox{[x . x . . x . x .]},
it is found in Bulgaria and Greece~\cite{arom-04}.
It is the rhythm of the
Macedonian dance \emph{Devoj\v{c}e}~\cite{singer-74}.
Finally, when started on the fourth onset, as in \rbox{[x . . x . x . x .]},
it is a rhythm found in Turkey~\cite{arom-04}, and is the metric
pattern of \emph{Strawberry Soup} by Don Ellis~\cite{keith-91}.

\mbox{$E(4,11) = $ \rbox{[x . . x . . x . . x .]} $ = (3332)$}
is the metric pattern
of the Dhruva t\={a}la of Southern India~\cite{london-04}.
It is also used by Frank Zappa in \emph{Outside Now}~\cite{keith-91}.
When it is started on the third onset, as in \rbox{[x . . x . x . . x . .]},
it is a Serbian rhythmic pattern~\cite{arom-04}.
When it is started on the fourth (last) onset it is the
\emph{Daasa al kbiri} rhythmic pattern of Yemen~\cite{hagoel-03}.

\mbox{$E(4,15) = $ \rbox{[x . . . x . . . x . . . x . .]} $ = (4443)$}
is the metric pattern
of the \emph{pa\~ncam sav\={a}r\={\i} t\={a}l}
of North Indian music~\cite{clayton-00}.

\mbox{$E(5,6) = $ \rbox{[x x x x x .]} $ = (11112)$}
yields the \emph{York-Samai} pattern, a popular
Arabic rhythm \cite{standifer-88}.
It is also a handclapping
rhythm used in the \emph{Al Med\={e}mi} songs of Oman~\cite{el-mallah-90}.

\mbox{$E(5,7) = $ \rbox{[x . x x . x x]} $ = (21211)$}
is the \emph{Nawakhat} pattern,
another popular Arabic rhythm \cite{standifer-88}.
In Nubia it is called the \emph{Al Noht} rhythm~\cite{hagoel-03}.

\mbox{$E(5,8) = $ \rbox{[x . x x . x x .]} $ = (21212)$}
is the Cuban \emph{cinquillo} pattern discussed
in the preceding~\cite{floyd-99}, the \emph{Malfuf} rhythmic pattern
of Egypt~\cite{hagoel-03},
as well as the Korean \emph{Nong P'y\v{o}n}
drum pattern~\cite{hye-ku-81}.
Started on the second onset, it is a popular Middle Eastern
rhythm~\cite{wade-04}, as well as the \emph{Timini} rhythm
of Senegal, the \emph{Adzogbo} dance rhythm of Benin~\cite{chernoff-79},
the \emph{Spanish Tango}~\cite{evans-66}, the \emph{Maksum} of Egypt~\cite{hagoel-03},
and a 13th century Persian
rhythm, the \emph{Al-saghil-al-sani}~\cite{wright-78}.
When it is started on the third onset it is the \emph{M\"usemmen}
rhythm of Turkey~\cite{bektas-05}.
When it is started on the fourth onset it is the \emph{Kromanti}
rhythm of Surinam.

\mbox{$E(5,9) = $ \rbox{[x . x . x . x . x]} $ = (22221)$}
is a popular Arabic rhythm called
\emph{Agsag-Samai}~\cite{standifer-88}.
Started on the second onset,
it is a drum pattern used by the \emph{Venda} in South Africa~\cite{rahn-87},
as well as a Rumanian folk-dance rhythm~\cite{proca-69}.
It is also the rhythmic pattern of the \emph{Sigaktistos}
rhythm of Greece~\cite{hagoel-03}, and the \emph{Samai aktsak}
rhythm of Turkey~\cite{hagoel-03}.
Started on the third onset, it is the rhythmic pattern
of the \emph{Nawahiid} rhythm of Turkey~\cite{hagoel-03}.

\mbox{$E(5,11) = $ \rbox{[x . x . x . x . x . .]} $ = (22223)$}
is the metric pattern of the
Sav\={a}r\={\i} t\={a}la used in the Hindustani music of India~\cite{london-04}.
It is also a rhythmic pattern used in Bulgaria and Serbia~\cite{arom-04}.
In Bulgaria is is used in the \emph{Kopanitsa}~\cite{rice-04}.
This metric pattern has been used by Moussorgsky in
\emph{Pictures at an Exhibition}~\cite{keith-91}.
Started on the third onset, it is the rhythm of the
Macedonian dance \emph{Kalajdzijsko Oro}~\cite{singer-74},
and it appears in Bulgarian music as well~\cite{arom-04}.

\mbox{$E(5,12) = $ \rbox{[x . . x . x . . x . x .]} $ = (32322)$}
is a common rhythm played in the Central African Republic
by the \emph{Aka} Pygmies~\cite{arom-91, chemillier-02, chemillier-03}.
It is also the \emph{Venda} clapping
pattern of a South African children's song~\cite{pressing-83},
and a rhythm pattern used in Macedonia~\cite{arom-04}.
Started on the second onset, it is the \emph{Columbia} bell pattern
popular in Cuba and West Africa~\cite{klower-97}, as well as
a drumming pattern used in the \emph{Chakacha} dance of Kenya~\cite{barz-04}.
and also used in Macedonia~\cite{arom-04}.
Started on the third onset, it is the \emph{Bemba} bell pattern
used in Northern Zimbabwe~\cite{pressing-83}, and the rhythm of the
Macedonian dance \emph{Ibraim Od\v{z}a Oro}~\cite{singer-74}.
Started on the fourth onset, it is the \emph{Fume Fume} bell pattern
popular in West Africa~\cite{klower-97}, and is a rhythm used
in the former Yugoslavia~\cite{arom-04}.
Finally, when started on the fifth onset it is the \emph{Salve} bell
pattern used in the Dominican Republic in a rhythm called \emph{Canto de Vela}
in honor of the Virgin Mary~\cite{farquharson-92}, as well as the drum
rhythmic pattern of the Moroccan \emph{Al Kud\'am}~\cite{hagoel-03}.

\mbox{$E(5,13) = $ \rbox{[x . . x . x . . x . x . .]} $ = (32323)$}
is a Macedonian
rhythm which is also played by starting it on the
fourth onset as follows: \rbox{[x . x . . x . . x . x . .]}~\cite{arom-04}.

\mbox{$E(5,16) = $ \rbox{[x . . x . . x . . x . . x . . .]} $ = (33334)$}
is the \emph{Bossa-Nova}
rhythm necklace of Brazil.
The actual Bossa-Nova rhythm usually starts on the third
onset as follows: \rbox{[x . . x . . x . . . x . . x . .]}~\cite{toussaint-02}.
However, other starting places are also documented in world music practices,
such as \rbox{[x . . x . . x . . x . . . x . .]}~\cite{behague-73}.

\mbox{$E(6,7) = $ \rbox{[x x x x x x .]} $ = (111112)$}
is the \emph{P\'ontakos} rhythm of Greece when started on the
sixth (last) onset~\cite{hagoel-03}.

\mbox{$E(6,13) = $ \rbox{[x . x . x . x . x . x . .]} $ = (222223)$}
is the rhythm of the
Macedonian dance \emph{Mama Cone pita}~\cite{singer-74}.
Started on the third onset, it is the rhythm of the
Macedonian dance \emph{Postupano Oro}~\cite{singer-74}, as well as
the \emph{Krivo Plovdivsko Horo} of Bulgaria~\cite{rice-04}.

\mbox{$E(7,8) = $ \rbox{[x x x x x x x .]} $ = (1111112)$}, when started on the
seventh (last) onset, is a typical rhythm played
on the \emph{Bendir} (frame drum), and used in the
accompaniment of songs of the \emph{Tuareg} people of Libya~\cite{standifer-88}.

\mbox{$E(7,9) = $ \rbox{[x . x x x . x x x]} $ = (2112111)$}
is the \emph{Bazaragana} rhythmic
pattern of Greece~\cite{hagoel-03}.

\mbox{$E(7,10) = $ \rbox{[x . x x . x x . x x]} $ = (2121211)$}
is the \emph{Lenk fahhte} rhythmic
pattern of Turkey~\cite{hagoel-03}.

\mbox{$E(7,12) = $ \rbox{[x . x x . x . x x . x .]} $ = (2122122)$}
is a common West African
bell pattern. For example, it is used in the \emph{Mpre} rhythm of
the \emph{Ashanti} people of Ghana~\cite{toussaint-03}.
Started on the seventh (last) onset, it is a \emph{Yoruba} bell pattern
of Nigeria, a \emph{Babenzele} pattern of Central Africa,
and a \emph{Mende} pattern of Sierra Leone~\cite{stone-05}.

\mbox{$E(7,15) = $ \rbox{[x . x . x . x . x . x . x . .]} $ = (2222223)$}
is a Bulgarian rhythm when started on the third onset~\cite{arom-04}.

\mbox{$E(7,16) = $ \rbox{[x . . x . x . x . . x . x . x .]} $ = (3223222)$}
is a \emph{Samba} rhythm necklace from Brazil. The actual Samba rhythm is
\rbox{[x . x . . x . x . x . . x . x .]} obtained by starting
$E(7,16)$ on the last onset, and it coincides with
a Macedonian rhythm~\cite{arom-04}.
When $E(7,16)$ is started on the fifth onset it is a clapping
pattern from Ghana~\cite{pressing-83}.
When it is started on the second onset it is a rhythmic pattern found
in the former Yugoslavia~\cite{arom-04}.

\mbox{$E(7,17) = $ \rbox{[x . . x . x . . x . x . . x . x .]} $ = (3232322)$}
is a Macedonian rhythm when started on the second onset~\cite{singer-74}.

\mbox{$E(7,18) = $ \rbox{[x . . x . x . . x . x . . x . x . .]} $ = (3232323)$}
is a Bulgarian rhythmic pattern~\cite{arom-04}.

\mbox{$E(8,17) = $ \rbox{[x . x . x . x . x . x . x . x . .]} $ = (22222223)$}
is a Bulgarian rhythmic pattern which is also started on the
fifth onset~\cite{arom-04}.

\mbox{$E(8,19) = $ \rbox{[x . . x . x . x . . x . x . x . . x .]} $ = (32232232)$}
is a Bulgarian rhythmic pattern when started on the
second onset~\cite{arom-04}.

\mbox{$E(9,14) = $ \rbox{[x . x x . x x . x x . x x .]} $ = (212121212)$},
when started on the
second onset, is the rhythmic pattern of the \emph{Tsofyan}
rhythm of Algeria~\cite{hagoel-03}.

\mbox{$E(9,16) = $ \rbox{[x . x x . x . x . x x . x . x .]} $ = (212221222)$}
is a rhythm necklace used in the Central African Republic~\cite{arom-91}.
When it is started on the second onset it is a bell pattern of
the \emph{Luba} people of Congo~\cite{putumayo-02}.
When it is started on the fourth onset it is a rhythm played in
West and Central Africa~\cite{floyd-99}, as well as a cow-bell pattern
in the Brazilian \emph{samba}~\cite{sole-96}.
When it is started on the penultimate onset it is the bell pattern
of the \emph{Ngbaka-Maibo} rhythms of the Central African Republic~\cite{arom-91}.

\mbox{$E(9,22) = $\rbox{[x . . x . x . . x . x . . x . x . . x . x .]} $ = (323232322)$}
is a Bulgarian rhythmic pattern when started on the
second onset~\cite{arom-04}.

\mbox{$E(9,23) = $\rbox{[x . . x . x . . x . x . . x . x . . x . x . .]}
$ = (323232323)$}
is a Bulgarian rhythm~\cite{arom-04}.

\mbox{$E(11,12) = $\rbox{[x x x x x x x x x x x .]} $ = (11111111112)$},
when started on the second onset, is the drum pattern
of the \emph{Rahm\={a}ni} (a cylindrical double-headed drum) used in the
\emph{S\={o}t sil\={a}m} dance from \emph{Mirb\={a}t} in the
South of Oman~\cite{el-mallah-90}.

\mbox{$E(11,24) = $\rbox{[x . . x . x . x . x . x . . x . x . x . x . x .]}
$ = (32222322222)$}
is a rhythm necklace of the \emph{Aka} Pygmies of Central Africa~\cite{arom-91}.
It is usually started on the seventh onset.
Started on the second onset, it is a Bulgarian rhythm~\cite{arom-04}.

\mbox{$E(13,24) = $\rbox{[x . x x . x . x . x . x . x x . x . x . x . x .]}
$ = (2122222122222)$}
is another rhythm necklace of the \emph{Aka} Pygmies of the upper
\emph{Sangha}~\cite{arom-91}.
Started on the penultimate onset, it is the \emph{Bobangi}
metal-blade pattern used by the \emph{Aka} Pygmies.

\mbox{$E(15,34) = $
\rbox{[x . . x . x . x . x . . x . x . x . x . . x . x . x . x . . x . x .]}
$ = (322232223222322)$}
is a Bulgarian rhythmic pattern when started on the
penultimate onset~\cite{arom-04}.

\subsection{Aksak Rhythms}

Euclidean rhythms are closely related to
a family of rhythms known as \emph{aksak}
rhythms, which have been studied from the combinatorial point
of view for some time~\cite{brailoiu-51, cler-94, arom-04}.
B\'ela Bart\'ok~\cite{bartok-81} and Constantin Br\u{a}iloiu~\cite{brailoiu-51},
respectively, have used the terms \emph{Bulgarian rhythm} and \emph{aksak}
to refer to those meters that use units of durations 2 and 3, and no other
durations. Furthermore, the rhythm or meter must contain at least one duration
of length 2 and at least one duration of length 3. Arom~\cite{arom-04}
refers to these durations as \emph{binary cells} and \emph{ternary cells},
respectively.

Arom~\cite{arom-04} generated an inventory of all the theoretically
possible \emph{aksak}
rhythms for values of $n$ ranging from 5 to 29, as well as a list of those
that are actually used in traditional world music.
He also proposed
a classification of these rhythms into several classes, based on
structural and numeric properties.
Three of his classes are considered here:
%
\begin{enumerate}
\item An \emph{aksak} rhythm is \emph{authentic}
      if $n$ is a \emph{prime} number.
\item An \emph{aksak} rhythm is \emph{quasi-aksak}
      if $n$ is an \emph{odd} number that is not prime.
\item An \emph{aksak} rhythm is \emph{pseudo-aksak}
      if $n$ is an \emph{even} number.
\end{enumerate}

A quick perusal of the Euclidean rhythms listed in the preceding
reveals that \emph{aksak} rhythms are well represented.  Indeed, all
three of Arom's classes (authentic, quasi-aksak, and pseudo-aksak)
make their appearance. There is a simple characterization of those
Euclidean rhythms that are \emph{aksak}.
>From the iterative subtraction algorithm of Bjorklund it follows that
if $n=2k$ all cells are binary (duration 2).
Similarly, if $n=3k$ all cells are ternary (duration 3).
Therefore, to ensure that the Euclidean rhythm contains both binary
and ternary cells, and no other durations, it follows that
$n$ must be between $2k$ and $3k$. 

Of course, not all \emph{aksak} rhythms are Euclidean.
Consider the Bulgarian rhythm with interval sequence (3322)~\cite{arom-04},
which is also the metric pattern of \emph{Indian Lady} by Don Ellis~\cite{keith-91}.
Here $k=4$ and $n=10$, and \mbox{$E(4,10) = $ \rbox{[x . . x . x . . x .]}} or
$(3232)$, a periodic rhythm.

\medskip

The following Euclidean rhythms are \emph{authentic aksak}:

\medskip

\noindent \mbox{$E(2,5) = $ \rbox{[x . x . .]} $ =  (23)$}
(classical music, jazz, Greece, Macedonia, Namibia, Persia, Rwanda).

\noindent \mbox{$E(3,7) = $ \rbox{[x . x . x . .]} $ =  (223)$}
(Bulgaria, Greece, Sudan, Turkestan).

\noindent \mbox{$E(4,11) = $ \rbox{[x . . x . . x . . x .]} $ =  (3332)$}
(Southern India rhythm), (Serbian necklace).

\noindent \mbox{$E(5,11) = $ \rbox{[x . x . x . x . x . .]} $ =  (22223)$}
(classical music, Bulgaria, Northern India, Serbia).

\noindent \mbox{$E(5,13) = $ \rbox{[x . . x . x . . x . x . .]} $ =  (32323)$} (Macedonia).

\noindent \mbox{$E(6,13) = $ \rbox{[x . x . x . x . x . x . .]} $ =  (222223)$} (Macedonia).

\noindent \mbox{$E(7,17) = $ \rbox{[x . . x . x . . x . x . . x . x .]} $ =  (3232322)$}
(Macedonian necklace).

\noindent \mbox{$E(8,17) = $  \rbox{[x . x . x . x . x . x . x . x . .]} $ =  (22222223)$}
(Bulgaria).

\noindent \mbox{$E(8,19) = $ \rbox{[x . . x . x . x . . x . x . x . . x .]} $ =  (32232232)$}
(Bulgaria).

\noindent \mbox{$E(9,23) = $ \rbox{[x . . x . x . . x . x . . x . x . . x . x . .]} $ =  (323232323)$}
(Bulgaria).

\medskip

The following Euclidean rhythms are \emph{quasi-aksak}:

\medskip

\noindent \mbox{$E(4,9) = $ \rbox{[x . x . x . x . .]} $ =  (2223)$} (Greece, Macedonia,
Turkey, Za\"{\i}re).

\noindent \mbox{$E(7,15) = $ \rbox{[x . x . x . x . x . x . x . .]} $ =  (2222223)$}
(Bulgarian necklace).

\medskip

The following Euclidean rhythms are \emph{pseudo-aksak}:

\medskip

\noindent \mbox{$E(3,8) = $ \rbox{[x . . x . . x .]} $ =  (332)$}
(Central Africa, Greece, India, Latin America, West Africa, Sudan).

\noindent \mbox{$E(5,12) = $ \rbox{[x . . x . x . . x . x .]} $ =  (32322)$}
(Macedonia, South Africa).

\noindent \mbox{$E(7,16) = $ \rbox{[x . . x . x . x . . x . x . x .]} $ =  (3223222)$}
(Brazilian, Macedonian, West African necklaces).

\noindent \mbox{$E(7,18) = $ \rbox{[x . . x . x . . x . x . . x . x . .]} $ =  (3232323)$}
(Bulgaria).

\noindent \mbox{$E(9,22) = $ \rbox{[x . . x . x . . x . x . . x . x . . x . x .]} $ =  (323232322)$}
(Bulgarian necklace).

\noindent \mbox{$E(11,24) = $ \rbox{[x . . x . x . x . x . x . . x . x . x . x . x .]} $ =  (32222322222)$}
(Central African and Bulgarian necklaces).

\noindent \mbox{$E(15,34) = $
\rbox{[x . . x . x . x . x . . x . x . x . x . . x . x . x . x . . x . x .]} $ =  (322232223222322)$}
(Bulgarian necklace).

\subsection{Drawing Digital Straight Lines}

\begin{figure}
  \centering
  \includegraphics*[scale=0.6]{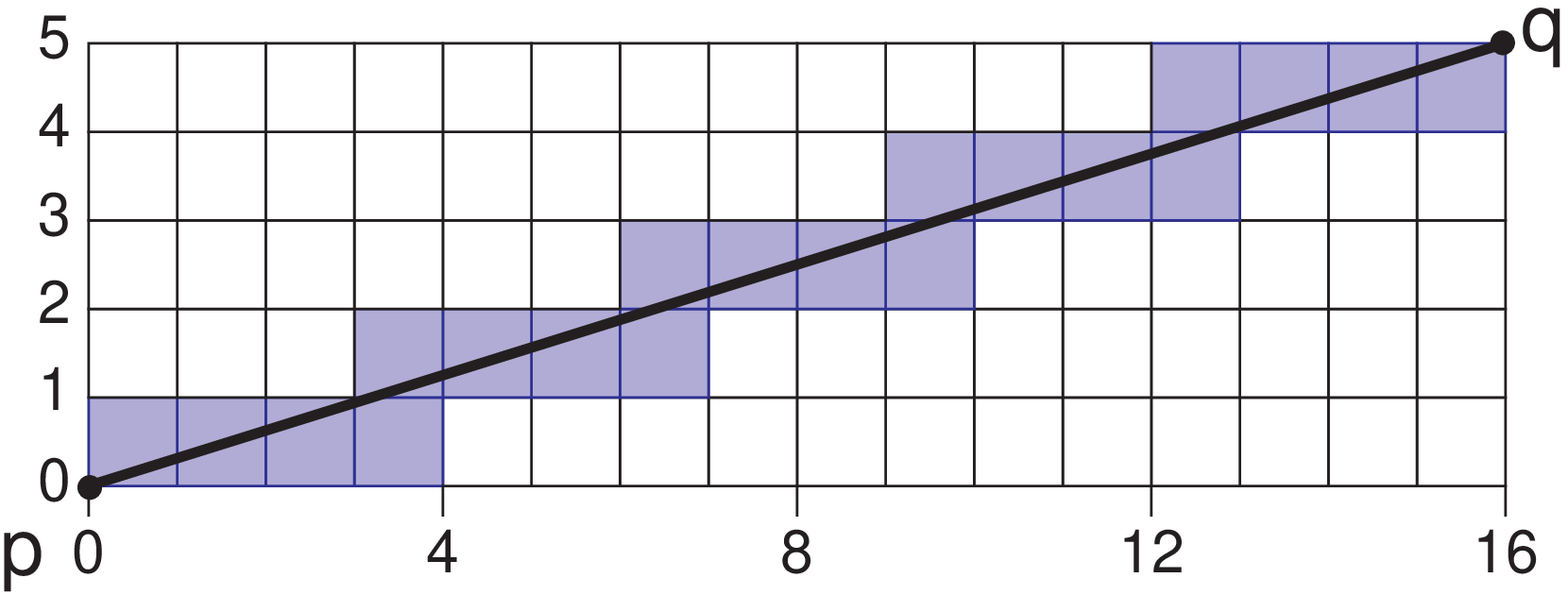}
  \caption{{The shaded pixels form a digital straight line determined by the
            points $p$ and $q$.}}
  \label{digital-straight-line}
\end{figure}

Euclidean rhythms and necklace patterns
also appear in the computer graphics literature on
drawing digital straight lines~\cite{klette-04}.
The problem here consists of efficiently converting a mathematical
straight line segment defined by the $x$ and $y$ integer coordinates
of its endpoints, to an ordered sequence of pixels that most faithfully
represents the given straight line segment.
Figure~\ref{digital-straight-line} illustrates an example of a digital
straight line (shaded pixels) determined by the two given endpoints
$p$ and $q$.
All the pixels intersected by the segment $(p,q)$ are shaded.
If we follow either the lower or upper boundary of the shaded pixels
from left to right we obtain the interval sequences
(43333) or (33334), respectively.
Note that the upper pattern corresponds to $E(5,16)$, a \emph{Bossa-Nova} variant.
Indeed, Harris and Reingold~\cite{harris-04} show that the well-known
Bresenham algorithm~\cite{bresenham-65} is described by the Euclidean algorithm.

\subsection{Calculating Leap Years in Calendar Design}

For thousands of years human beings have observed and measured the time it takes
between two consecutive sunrises, and between two consecutive spring seasons.
These measurements inspired different cultures to design 
calendars~\cite{ascher-02, reingold-01}.
Let $T_y$ denote the duration of one revolution of the earth around the sun,
more commonly known as a year.
Let $T_d$ denote the duration of one complete rotation of the earth,
more commonly known as a day.
The values of $T_y$ and $T_d$ are of course continually changing, because
the universe is continually reconfiguring itself.
However the ratio $T_y/T_d$ is approximately 365.242199.....
It is very convenient therefore to make a year last 365 days.
The problem that arises both for history and for predictions of the future,
is that after a while the 0.242199..... starts to contribute to a large error.
One simple solution is to add one extra day every 4 years: the so-called
Julian calendar. A day with one extra day is called a leap year.
But this assumes that a year is 365.25 days long, which is still
slightly greater than 365.242199......
So now we have an error in the opposite direction albeit smaller.
One solution to this problem is the Gregorian calendar~\cite{shallit-94}.
The Gregorian calendar defines a leap year as one divisible by 4,
except not those divisible by 100, except not those divisible by 400.
With this rule a year becomes 365 + 1/4 - 1/100 + 1/400 = 365.2425
days long, not a bad approximation.

Another solution is provided by the Jewish calendar which uses the idea
of cycles~\cite{ascher-02}.
Here a regular year has 12 months and a leap year
has 13 months. The cycle has 19 years including 7 leap years.
The 7 leap years must be distributed as evenly as possible
in the cycle of 19.
The cycle is assumed to start with Creation as year 1.
If the year modulo 19 is one of 3, 6, 8, 11, 14, 17, or 19,
then it is a leap year.
For example, the year $5765 = 303 \cdot 19 + 8$ and so is a leap year.
The year 5766, which begins at
sundown on the Gregorian date of October 3, 2005, is 5766 = 303x19 + 9,
and is therefore not a leap year.
Applying Bjorklund's algorithm to the integers 7 and 19 yields
\mbox{$E(7,19) = $ \rbox{[x . . x . x . . x . . x . x . . x . .]}}.
If we start this rhythm at the 7th pulse we obtain the pattern
\rbox{[. . x . . x . x . . x . . x . . x . x]}, which describes
precisely the leap year pattern 3, 6, 8, 11, 14, 17, and 19
of the Jewish calendar.
In this sense the Jewish calendar is an instance of a Euclidean necklace.

\subsection{Euclidean Strings}

In the study of the combinatorics of words and sequences, there exists
a family of strings called Euclidean strings~\cite{ellis-03}.
In this section we explore the relationship between Euclidean
strings and Euclidean rhythms.
We use the same terminology and notation introduced in~\cite{ellis-03}.

\begin{figure}
  \centering
  \includegraphics*[scale=0.7]{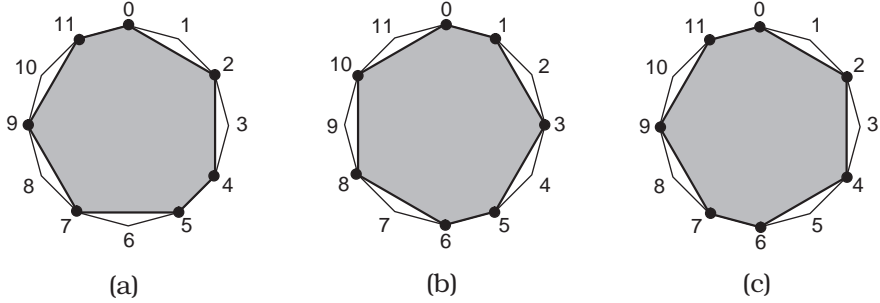}
  \caption{{Two right-rotations of the \emph{Bemb\'{e}} string:
           (a) the \emph{Bemb\'{e}}, (b) rotation by one unit, (c) rotation by seven units.}}
  \label{bembe.rotations}
\end{figure}

Let $P = (p_0,p_1,...,p_{n-1})$ denote a string of non-negative integers.
Let $\rho(P)$ denote the right rotation of $P$ by one position; that is,
$\rho(P) = (p_{n-1},p_0,p_1,...,p_{n-2})$.
Let $\rho^d(P)$ denote the right rotation of $P$ by $d$ positions.
If $P$ is considered as a cyclic string, a right rotation corresponds
to a clockwise rotation.
Figure~\ref{bembe.rotations} illustrates the $\rho(P)$ operator
with $P$ equal to the \emph{Bemb\'{e}} bell-pattern of
West Africa~\cite{toussaint-03}.
Figure~\ref{bembe.rotations}(a) shows the \emph{Bemb\'{e}} bell-pattern,
Figure~\ref{bembe.rotations}(b) shows $\rho(P)$,
which is a hand-clapping pattern from West Africa~\cite{pressing-83}, and
Figure~\ref{bembe.rotations}(c) shows $\rho^7(P)$,
which is the \emph{Tamb\'{u}} rhythm of Cura\c{c}ao~\cite{rosalia-02}.

Ellis et al.~\cite{ellis-03} define a string $P = (p_0,p_1,...,p_{n-1})$
to be \emph{Euclidean} if incrementing $p_0$ by~$1$ and decrementing
$p_{n-1}$ by~$1$ yields a new string $\tau(P)$ that is the rotation of $P$.
In other words, $P$ and  $\tau(P)$ are instances of the same  necklace.
Therefore, if we represent rhythms as binary sequences, Euclidean rhythms
cannot be Euclidean strings because 
all Euclidean rhythms begin with a `one'. Increasing $p_0$ by one
makes it a `two', which is not a binary string.
Therefore, to explore the relationship between Euclidean strings and
Euclidean rhythms, we will represent rhythms
by their clockwise distance sequences,
which are also strings of nonnegative integers.
As an example, consider \mbox{$E(4,9) = $ \rbox{[x . x . x . x . .]}} $= (2223)$.
Now \mbox{$\tau(2223) = (3222)$}, which is a
rotation of $E(4,9)$, and thus $(2223)$ is a Euclidean string.
Indeed, for $P=E(4,9)$, $\tau(P) = \rho^3(P)$.
As a second example, consider the West African clapping-pattern
shown in Figure~\ref{bembe.rotations}(b) given by $P=\mbox{(1221222)}$.
We have that $\tau(P) = \mbox{(2221221)} = \rho^6(P)$, the pattern
shown in Figure~\ref{bembe.rotations}(c), which also happens to be
the mirror image of $P$ about the $(0,6)$ axis.
Therefore $P$ is a Euclidean string.
However, note that $P$ is not a Euclidean rhythm.
Nevertheless, $P$ is a rotation of the Euclidean
rhythm \mbox{$E(7,12)=(2122122)$}.

Ellis et al.~\cite{ellis-03} have many beautiful results about
Euclidean strings. They show that Euclidean strings exist if,
and only if, $n$ and $(p_0+p_1+...+p_{n-1})$ are relatively prime
numbers, and that when they exist they are unique.
They also show how to construct Euclidean strings using an algorithm
that has the same structure as the Euclidean algorithm.
In addition they relate Euclidean strings to many other families of
sequences studied in the combinatorics of words~\cite{allouche-02, lothaire-02}.

Let $R(P)$ denote the reversal (or mirror image) of $P$; that is, 
$R(P)=(p_{n-1},p_{n-2},...,p_1,p_0)$.
Now we may determine which of the Euclidean rhythms used in
world music listed in the preceding, are Euclidean strings
or \emph{reverse} Euclidean strings.
The length of a Euclidean string is defined as the number of integers
it has. This translates in the rhythm domain to the number of onsets
a rhythm contains.
Furthermore, strings of length one are Euclidean strings, trivially.
Therefore all the trivial Euclidean rhythms with only one onset, such as
\mbox{$E(1,2) = $ \rbox{[x .]} $ = (2)$}, \mbox{$E(1,3) = $ \rbox{[x . .]} $ = (3)$},
and \mbox{$E(1,4) = $ \rbox{[x . . .]} $ = (4)$}, etc., are both Euclidean strings as
well as reverse Euclidean strings.
In the lists that follow the Euclidean rhythms are shown
in their box-notation format as well as in the clockwise distance sequence  
representation. The styles of music that use these rhythms is
also included. Finally, if only a rotated version of the Euclidean rhythm
is played, then it is still included in the list but referred to as a necklace.

\medskip

The following Euclidean rhythms are Euclidean strings:

\medskip

\noindent \mbox{$E(2,3) = $ \rbox{[x x .]} $ = (12)$}
(West Africa, Latin America, Nubia, Northern Canada).

\noindent \mbox{$E(2,5) = $ \rbox{[x . x . .]} $ = (23)$}
(classical music, jazz, Greece, Macedonia, Namibia, Persia, Rwanda), (\emph{authentic aksak}).

\noindent \mbox{$E(3,4) = $ \rbox{[x x x .]} $ = (112)$} (Brazil, Bali rhythms),
(Colombia, Greece, Spain, Persia, Trinidad necklaces).

\noindent \mbox{$E(3,7) = $ \rbox{[x . x . x . .]} $ = (223)$}
(Bulgaria, Greece, Sudan, Turkestan), (\emph{authentic aksak}).

\noindent \mbox{$E(4,5) = $ \rbox{[x x x x .]} $ = (1112)$} (Greece).

\noindent \mbox{$E(4,9) = $ \rbox{[x . x . x . x . .]} $ = (2223)$} (Greece, Macedonia,
Turkey, Za\"{\i}re), (\emph{quasi-aksak}).

\noindent \mbox{$E(5,6) = $ \rbox{[x x x x x .]} $ = (11112)$} (Arab).

\noindent \mbox{$E(5,11) = $ \rbox{[x . x . x . x . x . .]} $ = (22223)$}
(classical music, Bulgaria, Northern India, Serbia), (\emph{authentic aksak}).

\noindent \mbox{$E(5,16) = $ \rbox{[x . . x . . x . . x . . x . . . .]} $ = (33334)$}
(Brazilian, West African necklaces).

\noindent \mbox{$E(6,7) = $ \rbox{[x x x x x x .]} $ = (111112)$} (Greek necklace)

\noindent \mbox{$E(6,13) = $ \rbox{[x . x . x . x . x . x . .]} $ = (222223)$}
(Macedonia), (\emph{authentic aksak}).

\noindent \mbox{$E(7,8) = $ \rbox{[x x x x x x x .]} $ = (1111112)$} (Libyan necklace).

\noindent \mbox{$E(7,15) = $ \rbox{[x . x . x . x . x . x . x . .]} $ = (2222223)$} 
(Bulgarian necklace), (\emph{quasi-aksak}).

\noindent \mbox{$E(8,17) = $ \rbox{[x . x . x . x . x . x . x . x . .]} $ = (22222223)$}
(Bulgaria), (\emph{authentic aksak}).

\medskip

The following Euclidean rhythms are reverse Euclidean strings:

\medskip

\noindent \mbox{$E(3,5) = $ \rbox{[x . x . x]} $ = (221) $} (Korean, Rumanian, Persian necklaces).

\noindent \mbox{$E(3,8) = $ \rbox{[x . . x . . x .]} $ = (332)$}
(Central Africa, Greece, India, Latin America, West Africa, Sudan), (\emph{pseudo-aksak}).

\noindent \mbox{$E(3,11) = $ \rbox{[x . . . x . . . x . .]} $  = (443)$} (North India).

\noindent \mbox{$E(3,14) = $ \rbox{[x . . . . x . . . . x . . .]} $  = (554)$} (North India).

\noindent \mbox{$E(4,7) = $ \rbox{[x . x . x . x]} $  = (2221)$} (Bulgaria).

\noindent \mbox{$E(4,11) = $ \rbox{[x . . x . . x . . x .]} $  = (3332)$}
(Southern India rhythm), (Serbian necklace), (\emph{authentic aksak}).

\noindent \mbox{$E(4,15) = $ \rbox{[x . . . x . . . x . . . x . .]} $  = (4443)$} 
(North India).

\noindent \mbox{$E(5,7) = $ \rbox{[x . x x . x x]} $  = (21211)$} (Arab).

\noindent \mbox{$E(5,9) = $ \rbox{[x . x . x . x . x]} $  = (22221)$} (Arab).

\noindent \mbox{$E(5,12) = $ \rbox{[x . . x . x . . x . x .]} $  = (32322)$}
(Macedonia, South Africa), (\emph{pseudo-aksak}).

\noindent \mbox{$E(7,9) = $ \rbox{[x . x x x . x x x]} $  = (2112111)$} (Greece).

\noindent \mbox{$E(7,10) = $ \rbox{[x . x x . x x . x x]} $  = (2121211)$} (Turkey).

\noindent \mbox{$E(7,16) = $ \rbox{[x . . x . x . x . . x . x . x .]} $  = (3223222)$}
(Brazilian, Macedonian, West African necklaces), (\emph{pseudo-aksak}).

\noindent \mbox{$E(7,17) = $ \rbox{[x . . x . x . . x . x . . x . x .]} $  = (3232322)$}
(Macedonian necklace), (\emph{authentic aksak}).

\noindent \mbox{$E(9,22) = $ \rbox{[x . . x . x . . x . x . . x . x . . x . x .]} 
$  = (323232322)$}
(Bulgarian necklace), (\emph{pseudo-aksak}).

\noindent \mbox{$E(11,12) = $ \rbox{[x . x x x x x x x x x x]} $  = (11111111112)$} 
(Oman necklace).

\noindent \mbox{$E(11,24) = $ \rbox{[x . . x . x . x . x . x . . x . x . x . x . x .]} 
$  = (32222322222)$}
 (Central African and Bulgarian necklaces), (\emph{pseudo-aksak}).

\medskip

The following Euclidean rhythms are neither Euclidean nor reverse Euclidean
strings:

\medskip

\noindent \mbox{$E(5,8) = $ \rbox{[x . x x . x x .]} $  = (21212)$}
(Egypt, Korea, Latin America, West Africa).

\noindent \mbox{$E(5,13) = $ \rbox{[x . . x . x . . x . x . .]} $ = (32323)$} (Macedonia),
(\emph{authentic aksak}).

\noindent \mbox{$E(7,12) = $ \rbox{[x . x x . x . x x . x .]} $ = (2122122)$} (West Africa),
(Central African, Nigerian, Sierra Leone necklaces).

\noindent \mbox{$E(7,18) = $ \rbox{[x . . x . x . . x . x . . x . x . .]} $ = (3232323)$}
(Bulgaria), (\emph{pseudo-aksak}).

\noindent \mbox{$E(8,19) = $ \rbox{[x . . x . x . x . . x . x . x . . x .]} $ = (32232232)$}
(Bulgaria), (\emph{authentic aksak}).

\noindent \mbox{$E(9,14) = $ \rbox{[x . x x . x x . x x . x x .]} $ = (212121212)$} (Algerian necklace).

\noindent \mbox{$E(9,16) = $ \rbox{[x . x x . x . x . x x . x . x .]} $ = (212221222)$}
(West and Central African, and Brazilian necklaces).

\noindent \mbox{$E(9,23) = $ \rbox{[x . . x . x . . x . x . . x . x . . x . x . .]} 
$ = (323232323)$} (Bulgaria), (\emph{authentic aksak}).

\noindent \mbox{$E(13,24) = $ \rbox{[x . x x . x . x . x . x . x x . x . x . x . x .]} 
$ = (2122222122222)$} (Central African necklace).

\noindent \mbox{$E(15,34) =$
\rbox{[x . . x . x . x . x . . x . x . x . x . . x . x . x . x . . x . x .]} 
$ = (322232223222322)$} 
\linebreak
(Bulgarian necklace), (\emph{pseudo-aksak}).

\medskip

These three groups of Euclidean rhythms reveal a tantalizing pattern.
The Euclidean rhythms that are favored in classical music and jazz
are also Euclidean strings (the first group).
Furthermore, this group is not popular in African music.
The Euclidean rhythms that are neither Euclidean strings nor reverse
Euclidean strings (group three) fall into two categories:
those consisting of clockwise distances $1$ and $2$, and
those consisting of clockwise distances $2$ and $3$.
The latter group is used only in Bulgaria, and the former
is used in Africa.
Finally, the Euclidean rhythms that are reverse Euclidean strings
(the second group) appear to have a much wider appeal.
Finding musicological explanations for the preferences apparent
in these mathematical properties
raises interesting ethnomusicological questions.

The Euclidean strings defined in~\cite{ellis-03} determine
another family of rhythms, many of which are also used in world music
but are not necessarily Euclidean rhythms.
For example, \mbox{(1221222)} is an Afro-Cuban bell pattern.
Therefore it would be interesting to explore empirically the relation
between Euclidean strings and world music rhythms,
and to determine formally the exact mathematical relation between
Euclidean rhythms and Euclidean strings.

\section{Definitions and Notation}
\label{sec:Definitions}
\label{first technical section}

Before we begin the more technical part of the paper,
we need to define some precise mathematical notation for
describing rhythms.


Let $\mathbb{Z}^+$ denote the set of positive integers.
For $k,n \in \mathbb{Z}^+$, 
let $\gcd(k, n)$ denote the greatest common divisor of $k$ and~$n$.
If $\gcd(k,n) = 1$, we call $k$ and $n$ \emph{relatively prime}.
For integers $a < b$, let $[a,b] = \{a, a+1, a+2, \dots, b\}$. 

Let $C$ be a circle in the plane,
and consider any two points $x, y$ on~$C$.
The \emph{chordal distance} between $x$ and $y$, denoted by $\chordd(x,y)$, 
is the length of the line segment $\overline{x y}$; that is, 
$\chordd(x,y)$ is the Euclidean distance between $x$ and~$y$.
The \emph{clockwise distance} from $x$ to $y$,
or of the ordered pair $(x, y)$, is the length of the 
clockwise arc of $C$ from $x$ to $y$, and is denoted by $\clockwised(x,y)$.
Finally, the \emph{geodesic distance} between $x$ and~$y$, 
denoted by $\geodesicd(x,y)$, is the length of the shortest arc of $C$
between $x$ and $y$; that is, 
$\geodesicd(x,y) = \min\{\clockwised(x,y), \clockwised(y,x)\}$.

A \emph{rhythm of timespan $n$} is a subset of 
$\{0, 1, \dots, n-1\}$, representing the set of pulses that are onsets
in each repetition.
For clarity, we write the timespan $n$ as a subscript after the subset:
$\{\dots\}_n$.
Geometrically, if we locate $n$ equally spaced points clockwise
around a circle $C_n$ of circumference~$n$,
then we can view a rhythm of timespan $n$ as a subset of these $n$ points.
We consider an element of $C_n$ to simultaneously be a point on the 
circle and an integer in $\{0, 1, \dots, n-1\}$.

The \emph{rotation} of a rhythm $R$ of timespan $n$ by an integer $\Delta \geq 0$
is the rhythm $\{(i + \Delta) \bmod n : i \in R\}_n$ of the same timespan~$n$.
The \emph{scaling} of a rhythm $R$ of timespan $n$
by an integer $\alpha \geq 1$ is the rhythm $\{\alpha i : i \in R\}_{\alpha n}$
of timespan $\alpha n$.

Let $R = \{r_0,r_1, \dots, r_{k-1}\}_n$ be a rhythm of timespan $n$ 
with $k$ onsets sorted in clockwise order.
%
Throughout this paper, an onset $r_i$ will mean $(r_{i\bmod{k}})\bmod{n}$. 
Observe that the clockwise distance $\clockwised(r_i,r_j)=(r_j-r_i)\bmod{n}$. 
This is the number of points on $C_n$ that are contained in the clockwise arc $(r_i,r_j]$ 
and is also known as the \emph{chromatic length}~\cite{clough-91}.

%
%


The \emph{geodesic distance multiset} of a rhythm $R$ is the multiset
of all nonzero pairwise geodesic distances; that is, it is the multiset 
 $\{\geodesicd(r_i,r_j) : r_i,r_j \in R, r_i \neq r_j\}$.
The geodesic distance multiset has cardinality ${k \choose 2}$.
The \emph{multiplicity} of a distance $d$ is the number of occurrences
of $d$ in the geodesic distance multiset.

A rhythm is \emph{Erd\H{o}s-deep} if it has (exactly) one distance of
multiplicity~$i$, for each $i \in [1,k-1]$.
Note that these multiplicities sum to $\sum_{i=1}^{k-1} i = {k \choose 2}$, 
which is the cardinality of the geodesic distance multiset, and
hence these distances are all the distances in the rhythm.
Every geodesic distance is between $0$ and $\lfloor n/2 \rfloor$.
A rhythm is \emph{Winograd-deep} if
every two distances from $\{1, 2, \dots, \lfloor \frac{n}{2} \rfloor\}$
have different multiplicity.

A \emph{shelling} of an Erd\H{o}s-deep rhythm $R$ is an ordering
$s_1, s_2, \dots, s_k$ of the onsets in $R$ such that
$R - \{s_1, s_2, \dots, s_i\}$ is an Erd\H{o}s-deep rhythm
for $i = 0, 1, \dots, k$.
(Every rhythm with at most two onsets is Erd\H{o}s-deep.)


%
%
%

The \emph{evenness} of rhythm $R$ is the sum of all inter-onset chordal distances in $R$; 
that is, $\displaystyle\sum_{0\leq i<j\leq k-1}\chordd(r_i,r_j).$

The \emph{clockwise distance sequence} of $R$ is the circular sequence
$(d_0,d_1,\dots,d_{k-1})$ where $d_i=\clockwised(r_i,r_{i+1})$ for all $i \in [0,k-1]$. 
Observe that each $d_i \in \mathbb{Z}^+$ and $\sum_id_i=n$. 

\begin{observation}
  There is a one-to-one relationship between rhythms with $k$ onsets and
  timespan~$n$ and circular sequences $(d_0,d_1,\dots,d_{k-1})$ 
  where each $d_i \in \mathbb{Z}^+$ and $\sum_i d_i=n$.
\end{observation}

\section{Even Rhythms}
\label{sec:Even}

In this section we first describe three algorithms that generate even rhythms.
We then characterize rhythms with maximum evenness and show
that, for given numbers of pulses and onsets, the three described algorithms
generate the unique rhythm with maximum evenness.
As mentioned in the introduction, the measure of evenness considered here
is the pairwise sum of chordal distances.

The even rhythms characterized in this section were studied by Clough  
and Meyerson \cite{clough-85, clough-86} for the case where the numbers of
pulses and onsets are relatively prime.
This was subsequently expanded upon by Clough and  
Douthett~\cite{clough-91}.  
We revisit these results and provide an additional connection to  
rhythms (and scales) that are obtained from the Euclidean algorithm.
Most of these results are stated in~\cite{clough-91}. 
However our proofs are new, and in many cases are much more streamlined.

\subsection{Characterization}

We first present three algorithms for computing a rhythm with $k$ onsets,
timespan~$n$, for any $k\leq n$, that possess large evenness.

The first algorithm is by Clough and Douthett~\cite{clough-91}:
%
\begin{algorithmbox}{\textsc{Clough-Douthett}$(n,k)$}
\begin{enumerate*}
\item {\bf return} $\{\FLOOR{\frac{i n}{k}}:i\in[0,k-1]\}$
\end{enumerate*}
\end{algorithmbox}
%
\noindent
Because $k\leq n$, the rhythm output by \textsc{Clough-Douthett}$(n,k)$
has $k$ onsets as desired. 

The second algorithm is a geometric heuristic implicit in 
the work of Clough and Douthett~\cite{clough-91}:
%
\begin{algorithmbox}{\textsc{Snap}$(n,k)$}
\begin{enumerate*}
\item
  Let $D$ be a set of $k$ evenly spaced points on $C_n$
  such that $D \cap C_n=\emptyset$.
\item
  For each point $x\in D$,
  let $x'$ be the first point in $C_n$ clockwise from~$x$.
\item
  {\bf return} $\{x' : x \in D\}$
\end{enumerate*}
\end{algorithmbox}
%
\noindent
Because $k\leq n$, the clockwise distance between consecutive points in $D$
in the execution of \textsc{Snap}$(n,k)$ is at least that of
consecutive points in~$C_n$. 
Thus, $x'\ne y'$ for distinct $x,y\in D$,
so \textsc{Snap} returns a rhythm with $k$ onsets as desired.

The third algorithm is a recursive algorithm in the same mold as Euclid's
algorithm for greatest common divisors.
The algorithm uses the clockwise distance sequence notation
described in the introduction.
The resulting rhythm always defines the same necklace as
the Euclidean rhythms from Section~\ref{Euclidean rhythms};
that is, the only difference is a possible rotation.

\begin{algorithmbox}{\textsc{Euclidean}$(n,k)$}
\begin{enumerate*}
\item {\bf if} $k$ evenly divides $n$ {\bf then return}
  $\displaystyle(\underbrace{\tfrac{n}{k},\tfrac{n}{k},\dots,\tfrac{n}{k}}_k)$
\item $a \gets n \bmod k$
\item $(x_1,x_2,\dots,x_a) \gets $ \textsc{Euclidean}$(k,a)$
\item {\bf return}
  $\displaystyle (\underbrace{\floor{\tfrac{n}{k}},\dots,\floor{\tfrac{n}{k}}}_{x_1-1},
  \ceil{\tfrac{n}{k}}; \,
  \underbrace{\floor{\tfrac{n}{k}},\dots,\floor{\tfrac{n}{k}}}_{x_2-1},\ceil{\tfrac{n}{k}}; \,
  \dots; \,
  \underbrace{\floor{\tfrac{n}{k}},\dots,\floor{\tfrac{n}{k}}}_{x_a-1},\ceil{\tfrac{n}{k}})$
\end{enumerate*}
\end{algorithmbox}

As a simple example, consider $k=5$ and $n=13$.
The sequence of calls to \textsc{Euclidean}$(n,k)$ follows the same
pattern as the \textsc{Euclid} algorithm for greatest common divisors from
Section~\ref{Euclid gcd}, except that it now stops one step earlier:
$(13,5)$, $(5,3)$, $(3,2)$, $(2,1)$.
\xxx[Erik]{Reversing the argument order would be nice
           for consistency with \textsc{Euclid}.}
At the base of the recursion, we have
\textsc{Euclidean}$(2,1) = (2) = $ \rbox{[x .]}.
At the next level up, we obtain
\textsc{Euclidean}$(3,2) = (1,2) = $ \rbox{[x x .]}.
Next we obtain
\textsc{Euclidean}$(5,3) = (2;1,2) = $ \rbox{[x . x x .]}.
Finally, we obtain
\textsc{Euclidean}$(13,5) = (2,3;3;2,3) = $ \rbox{[x . x . . x . . x . x . .]}.
(For comparison, the Euclidean rhythm from Section~\ref{Bjorklund algorithm}
is $E(5,13) = (2,3,2,3,3)$, a rotation by~$5$.)

We now show that algorithm \textsc{Euclidean}$(n,k)$ outputs a circular 
sequence of $k$ integers that sum to $n$ (which is thus the clockwise 
distance sequence of a rhythm with $k$ onsets and timespan~$n$). 
We proceed by induction on $k$.
If $k$ evenly divides $n$, then the claim clearly holds. 
Otherwise $a$ ($=n\bmod{k}$) $>0$, and by induction $\sum_{i=1}^ax_i=k$. 
Thus the sequence that is output has $k$ terms and sums to 
\begin{align*}
a\CEIL{\frac{n}{k}}+\FLOOR{\frac{n}{k}}\sum_{i=1}^a(x_i-1)
&=a\CEIL{\frac{n}{k}}+(k-a)\FLOOR{\frac{n}{k}}\\
&=a\bracket{1+\FLOOR{\frac{n}{k}}}+(k-a)\FLOOR{\frac{n}{k}}\\
&=a+k\FLOOR{\frac{n}{k}}\\
&=n\enspace.
\end{align*}


The following theorem is one of the main contributions of this paper. 

\begin{theorem}
\label{Main_Even}

Let $n\geq k\geq2$ be integers.
The following are equivalent for a rhythm $R=\{r_0,r_1,\dots,r_{k-1}\}_n$
with $k$ onsets and timespan~$n$:
\begin{enumerate}
\item[\AAA] $R$ has maximum evenness
            (sum of pairwise inter-onset chordal distances),
\item[\BBB] $R$ is a rotation of the \textsc{Clough-Douthett}$(n,k)$ rhythm,
\item[\CCC] $R$ is a rotation of the \textsc{Snap}$(n,k)$ rhythm,
\item[\DDD] $R$ is a rotation of the \textsc{Euclidean}$(n,k)$ rhythm,
\item[\STAR] for all $\ell\in[1,k]$ and $i\in[0,k-1]$,
    the ordered pair $(r_i,r_{i+\ell})$ has clockwise distance
    $\clockwised(r_i,r_{i+\ell})\in\{\floor{\frac{\ell n}{k}},\ceil{\frac{\ell n}{k}}\}$.
\end{enumerate}
Moreover, up to a rotation, there is a unique rhythm that satisfies
these conditions.
\end{theorem}

Note that the evenness of a rhythm equals the evenness of the same rhythm played backwards.
Thus, if $R$ is the unique rhythm with maximum evenness, then $R$ is the same
rhythm as $R$ played backwards (up to a rotation).

The proof of Theorem~\ref{Main_Even} proceeds as follows. 
In Section~\ref{Algorith Properties} we prove that each of the three algorithms 
produces a rhythm that satisfies property \STAR. 
Then in Section~\ref{Uniqueness} we prove that there is a unique 
rhythm that satisfies property \STAR. 
Thus the three algorithms produce the same rhythm, up to rotation.
Finally in Section~\ref{Max_Evenness} we prove that the unique rhythm that 
satisfies property \STAR\ maximizes evenness.


\subsection{Properties of the Algorithms} 
\label{Algorith Properties}

We now prove that each of the algorithms has property \STAR. 
Clough and Douthett~\cite{clough-91} proved the following.

\begin{proof}[Proof \BBB\ $\Rightarrow$ \STAR] 
Say $R=\{r_0,r_1,\dots,r_{k-1}\}_n$ is the \textsc{Clough-Douthett}$(n,k)$ rhythm. 
Consider an ordered pair $(r_i,r_{i+\ell})$ of onsets in $R$. 
Let $p_i = i n\bmod{k}$ and let $p_\ell = \ell n \bmod k$. 
By symmetry we can suppose that $r_i\leq r_{(i+\ell)\bmod{k}}$. 
Then the clockwise distance $\clockwised(r_i,r_{i+\ell})$ is
\begin{equation*}
\FLOOR{\frac{(i+\ell)n}{k}}-\FLOOR{\frac{in}{k}}
=
\FLOOR{\frac{in}{k}}+\FLOOR{\frac{\ell n}{k}}+
\FLOOR{\frac{p_i+p_\ell}{k}}-\FLOOR{\frac{in}{k}}
=
\FLOOR{\frac{\ell n}{k}}+\FLOOR{\frac{p_i+p_\ell}{k}}\enspace,
\end{equation*}
which is \FLOOR{\frac{\ell n}{k}} or
\CEIL{\frac{\ell n}{k}}, because $\FLOOR{\frac{p_i+p_\ell}{k}}\in\{0,1\}$.
\end{proof}

A similar proof shows that the rhythm $\{\CEIL{\frac{in}{k}}:i\in[0,k-1]\}$ 
satisfies property \STAR. Observe that \STAR\ is equivalent to the following property.

\begin{itemize}
\item[\STARSTAR] if $(d_0,d_1,\dots,d_{k-1})$ is the clockwise distance sequence of $R$, then for all $\ell\in[1,k]$, the sum of any $\ell$ consecutive elements in $(d_0,d_1,\dots,d_{k-1})$ equals $\ceil{\frac{\ell n}{k}}$ or $\floor{\frac{\ell n}{k}}$.
\end{itemize}

\begin{proof}[Proof \CCC\ $\Rightarrow$ \STARSTAR]
Let $(d_0,d_1,\dots,d_{k-1})$ be the clockwise distance sequence of the 
rhythm determined by \textsc{Snap}$(n,k)$. 
For the sake of contradiction, suppose that for some $\ell\in[1,k]$, 
the sum of $\ell$ consecutive elements in $(d_0,d_1,\dots,d_{k-1})$ 
is greater than $\ceil{\frac{\ell n}{k}}$. The case in which the sum is 
less than $\floor{\frac{\ell n}{k}}$ is analogous. We can assume that 
these $\ell$ consecutive elements are $(d_0,d_1,\dots,d_{\ell-1})$.  
Using the notation defined in the statement of the algorithm, 
let $x_0,x_1,\dots,x_\ell$ be the points in $D$ such that $\clockwised(x'_i,x'_{i+1})=d_i$ 
for all $i\in[0,\ell-1]$. Thus $\clockwised(x'_1, x'_{\ell+1})\geq\ceil{\frac{\ell n}{k}}+1$. 
Now $\clockwised(x_{\ell+1},x'_{\ell+1})<1$. 
Thus $\clockwised(x'_1,x_{\ell+1})>\ceil{\frac{\ell n}{k}}\geq\frac{\ell n}{k}$, 
which implies that $\clockwised(x_1,x_{\ell+1})>\frac{\ell n}{k}$. 
This contradicts the fact that the points in $D$ were evenly spaced around $C_n$ 
in the first step of the algorithm.


\end{proof}


\begin{proof}[Proof \DDD\ $\Rightarrow$ \STARSTAR]
We proceed by induction on $k$. Let $R =\,$\textsc{Euclidean}$(n,k)$. 
If $k$ evenly divides~$n$,
then $R=(\frac{n}{k},\frac{n}{k},\dots,\frac{n}{k})$, which satisfies \DDD. 
Otherwise, let $a=n\bmod k$ and let $(x_1,x_2,\dots,x_a) =\,$\textsc{Euclidean}$(k,a)$. 
By induction, for all $\ell\in[1,a]$, the sum of any $\ell$ consecutive 
elements in $(x_1,x_2,\dots,x_a)$ equals $\floor{\frac{\ell k}{a}}$ or 
$\ceil{\frac{\ell k}{a}}$. 
Let $S$ be a sequence of $m$ consecutive elements in $R$. 
By construction, for some $1\leq i\leq j\leq a$, and for some $0\leq s\leq x_i-1$ 
and $0\leq t\leq x_j-1$, we have
\begin{equation*}
S=(\underbrace{\floor{\tfrac{n}{k}},\dots,\floor{\tfrac{n}{k}}}_s,\ceil{\tfrac{n}{k}},
\underbrace{\floor{\tfrac{n}{k}},\dots,\floor{\tfrac{n}{k}}}_{x_{i+1}-1},\ceil{\tfrac{n}{k}},
\dots,
\underbrace{\floor{\tfrac{n}{k}},\dots,\floor{\tfrac{n}{k}}}_{x_{j-1}-1},\ceil{\tfrac{n}{k}},
\underbrace{\floor{\tfrac{n}{k}},\dots,\floor{\tfrac{n}{k}}}_t)\enspace.
\end{equation*}
It remains to prove that $\FLOOR{\frac{mn}{k}}\leq\sum S\leq\CEIL{\frac{mn}{k}}$.

We first prove that $\sum S\geq\FLOOR{\frac{mn}{k}}$. 
We can assume the worst case for $\sum S$ to be minimal, 
which is when $s=x_i-1$ and $t=x_j-1$. Thus by induction,
\begin{equation*}
m+1
=\sum_{\alpha=i}^jx_\alpha
\leq
\CEIL{\frac{(j-i+1)k}{a}}
\enspace.
\end{equation*}
Hence
\begin{align*}
\frac{am}{k}
\leq
\frac{a}{k}\CEIL{\frac{(j-i+1)k}{a}}-\frac{a}{k}
\leq
\frac{a}{k}\bracket{\frac{(j-i+1)k+a-1}{a}}-\frac{a}{k}
=
j-i+1-\frac{1}{k}\enspace.
\end{align*}
Thus $\floor{\frac{am}{k}}\leq j-i$ and
\begin{align*}
\sum S
=
m\FLOOR{\frac{n}{k}}+j-i
\geq 
m\FLOOR{\frac{n}{k}}+\FLOOR{\frac{am}{k}}
=
\FLOOR{m\FLOOR{\frac{n}{k}}+\frac{am}{k}}
=
\FLOOR{\frac{m}{k}\bracket{k\FLOOR{\frac{n}{k}}+a}}
=
\FLOOR{\frac{mn}{k}}
\enspace.
\end{align*}

Now we prove that $\sum S\leq\FLOOR{\frac{mn}{k}}$. 
We can assume the worst case for $\sum S$ to be maximal, 
which is when $s=0$ and $t=0$. Thus by induction,
\begin{equation*}
m-1
=\sum_{\alpha=i+1}^{j-1}x_\alpha
\geq
\FLOOR{\frac{(j-i-1)k}{a}}\enspace.
\end{equation*}
Hence
\begin{align*}
\frac{am}{k}
\geq
\frac{a}{k}\FLOOR{\frac{(j-i-1)k}{a}}+\frac{a}{k}
\geq
\frac{a}{k}\bracket{\frac{(j-i-1)k-a+1}{a}}+\frac{a}{k}
=
j-i-1+\frac{1}{k}
\enspace.
\end{align*}
Thus $\ceil{\frac{am}{k}}\geq j-i$ and
\begin{align*}
\sum S
=
m\FLOOR{\frac{n}{k}}+j-i
\leq 
m\FLOOR{\frac{n}{k}}+\CEIL{\frac{am}{k}}
=
\CEIL{m\FLOOR{\frac{n}{k}}+\frac{am}{k}}
=
\CEIL{\frac{m}{k}\bracket{k\FLOOR{\frac{n}{k}}+a}}
=
\CEIL{\frac{mn}{k}}
\enspace.
\end{align*}
\end{proof}

\subsection{Uniqueness}
\label{Uniqueness}

In this section we prove that there is a unique rhythm satisfying the
conditions in Theorem~\ref{Main_Even}.
The following well-known number-theoretic lemmas will be useful.
Two integers $x$ and $y$ are \emph{inverses} modulo $m$
if $x y \equiv1 \pmod{m}$.

\begin{lemma}[{\cite[page~55]{Stillwell}}]
  \label{lem:Inverse}
  An integer $x$ has an inverse modulo $m$ if and only if
  $x$ and $m$ are relatively prime.
  Moreover, if $x$ has an inverse modulo $m$,
  then it has an inverse $y \in [1,m-1]$.
\end{lemma}

\begin{lemma} \label{lem:NoSolution}
  If $x$ and $m$ are relatively prime,
  then $i x \not\equiv j x \pmod{m}$ for all distinct $i,j \in [0,m-1]$.
\end{lemma}

\begin{proof}
  Suppose that $i x \equiv j x \pmod{m}$ for some $i,j\in[0,m-1]$.
  By Lemma~\ref{lem:Inverse}, $x$ has an inverse modulo~$m$.
  Thus $i \equiv j \pmod{m}$,
  and $i=j$ because $i,j\in[0,m-1]$.
\end{proof}


\begin{lemma} \label{lem:Number}
For all relatively prime integers $n$ and $k$ with $2 \leq k \leq n$,
there is an integer $\ell \in [1,k-1]$ such that:
\begin{enumerate}
\item[\textup{(a)}] $\ell n \equiv 1\pmod{k}$,
\item[\textup{(b)}] $i\ell\not\equiv j\ell\pmod{k}$ for all distinct $i,j\in[0,k-1]$, and
\item[\textup{(c)}] $i\floor{\frac{\ell n}{k}} \not\equiv
j\floor{\frac{\ell n}{k}}\pmod{n}$ for all distinct $i,j\in[0,k-1]$.
\end{enumerate}
\end{lemma}

\begin{proof}
By Lemma~\ref{lem:Inverse} with $x=n$ and $m=k$, $n$ has an inverse $\ell$ modulo $k$.
This proves (a).
Thus $k$ and $\ell$ are relative prime by Lemma~\ref{lem:Inverse}
with $x=\ell$ and $m=k$.
Hence (b) follows from Lemma~\ref{lem:NoSolution}.
Let $t = \floor{\frac{\ell n}{k}}$.
Then $\ell n=kt+1$.
By Lemma~\ref{lem:NoSolution} with $m=n$ and $x=t$ (and because $k\leq n$),
to prove (c) it suffices to show that $t$ and $n$ are relatively prime.
Let $g = \gcd(t,n)$.
Thus $\ell\frac{n}{g}=k\frac{t}{g}+\frac{1}{g}$.
Because $\frac{n}{g}$ and $\frac{t}{g}$ are integers,
$\frac{1}{g}$ is an integer and $g=1$.
This proves (c).
\end{proof}


The following theorem is the main result of this section.

\begin{theorem}
\label{thm:Uniqueness}
For all integers $n$ and $k$ with $2 \leq k \leq n$,
there is a unique rhythm with $k$ onsets and timespan~$n$
that satisfies property~\STAR, up to a rotation.
\end{theorem}

\begin{proof}
Let $R=\{r_0,r_1,\dots,r_{k-1}\}_n$ be a $k$-onset rhythm that satisfies \STAR.
Recall that the index of an onset is taken modulo $k$, and that the value of
an onset is taken modulo $n$.
That is, $r_i=x$ means that $r_{i\bmod{k}}=x\bmod{n}$.

Let $g = \gcd(n,k)$.
We consider three cases for the value of $g$.

\textbf{Case 1.} $g=k$:
Because $R$ satisfies property \STAR\ for $\ell=1$,
every ordered pair $(r_i,r_{i+1})$ has clockwise distance $\frac{n}{k}$.
By a rotation of $R$ we can assume that $r_0=0$.
Thus $r_i=\frac{in}{k}$ for all $i\in[0,k-1]$.
Hence $R$ is uniquely determined in this case.

\textbf{Case 2.} $g=1$  (see Figure~\ref{fig:CaseTwo}):
By Lemma~\ref{lem:Number}(a), there is an integer $\ell\in[1,k-1]$
such that $\ell n\equiv1\pmod{k}$.
Thus $\ell n=(k-1)\floor{\frac{\ell n}{k}}+\ceil{\frac{\ell n}{k}}$.
Hence, of the $k$ ordered pairs $(r_i,r_{i+\ell})$ of onsets,
$k-1$ have clockwise distance \floor{\frac{\ell n}{k}}
and one has clockwise distance \ceil{\frac{\ell n}{k}}.
By a rotation of $R$ we can assume that $r_0=0$ and
$r_{k-\ell}=n-\ceil{\frac{\ell n}{k}}$.
Thus $r_{i\ell}=i\floor{\frac{\ell n}{k}}$ for all $i\in[0,k-1]$;
that is,
$r_{(i\ell)\bmod{k}}=(i\floor{\frac{\ell n}{k}})\bmod{n}$.
By Lemma~\ref{lem:Number}(b) and (c), 
this defines the $k$ distinct onsets of $R$.
Hence $R$ is uniquely determined in this case.

\begin{figure}[htbp]
\begin{center}
\includegraphics{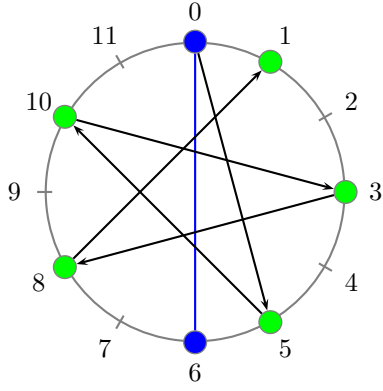}
\caption{
\label{fig:CaseTwo}
Here we illustrate Case 2 with $n=12$ and $k=7$.
Thus $\ell=3$ because $3\times 12\equiv1\pmod{7}$.
We have $\ceil{\tfrac{\ell n}{k}}=6$ and $\floor{\tfrac{\ell n}{k}}=5$.
By a rotation we can assume that $r_0=0$ and $r_{k-\ell}=r_4=6$ (the darker dots).
Then as shown by the arrows, the positions of the other onsets are implied.}
\end{center}
\end{figure}

\textbf{Case 3.} $g\in[2,k-1]$ (see Figure~\ref{fig:CaseThree}):
Let $k' = \frac{k}{g}$ and let $n' = \frac{n}{g}$.
Observe that both $k'$ and $n'$ are integers.
Because $R$ satisfies \STAR\ and
$\ceil{\frac{k'n}{k}}=\floor{\frac{k'n}{k}}=n'$,
we have $\clockwised(r_i,r_{i+k'})=n'$ for all $i\in[0,k-1]$.
Thus
\begin{equation}
\label{Blah}
r_{i k'+j}=i n'+r_j
\end{equation}
for all $i\in[0,g-1]$ and $j\in[0,n'-1]$.

Now $\gcd(n',k')=1$ by the maximality of $g$.
By Lemma~\ref{lem:Number}(a),
there is an integer $\ell'\in[1,k'-1]$ such that $\ell'n'\equiv1\pmod{k'}$.
Thus $\ell'n'=(k'-1)\floor{\frac{\ell'n'}{k'}}+\ceil{\frac{\ell'n'}{k'}}$,
implying $\ell'n=(k-g)\floor{\frac{\ell'n'}{k'}}+g\ceil{\frac{\ell'n'}{k'}}$.
Hence, of the $k$ ordered pairs $(r_i,r_{i+\ell'})$ of onsets,
$k-g$ have clockwise distance \floor{\frac{\ell'n'}{k'}}
and $g$ have clockwise distance \ceil{\frac{\ell'n'}{k'}}.
By a rotation of $R$ we can assume that $r_0=0$ and
$r_{\ell'}=\ceil{\frac{\ell'n'}{k'}}$.
By Equation~\eqref{Blah} with $j=0$ and $j=\ell'$, we have
\begin{equation}
\label{BlahBlah}
r_{i k'}=i n'\text{ and }r_{i k'+\ell'}=i n'+\CEIL{\tfrac{\ell'n'}{k'}}
\end{equation}
for all $i\in[0,g-1]$. This accounts for the $g$ ordered pairs
$(r_i,r_{i+\ell'})$ with clockwise distance \ceil{\frac{\ell'n'}{k'}}.
The other $k-g$ ordered pairs $(r_i,r_{i+\ell'})$ have clockwise
distance \floor{\frac{\ell'n'}{k'}}. Define
\begin{equation*}
L_0 = 0 \text{ and }
L_j = \CEIL{\tfrac{\ell'n'}{k'} }
+  (j-1)\FLOOR{\tfrac{\ell'n'}{k'} }
\text{ for all }  j  \in [ 1,k'-1]\enspace.
\end{equation*}
Thus by Equation~\eqref{BlahBlah},
\begin{equation*}
r_{i k'+j \ell'}=i n'+L_j
\end{equation*}
for all $i\in[0,g-1]$ and $j\in[0,k'-1]$;
that is, $r_{(i k'+j \ell')\bmod{k}}=(i n'+L_j)\bmod{n}$.

\begin{figure}[htbp]
\centering
\includegraphics{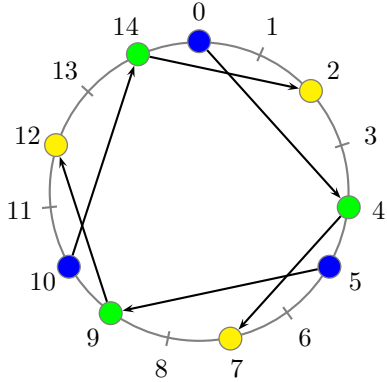}
\caption{
\label{fig:CaseThree}
Here we illustrate Case 3 with $n=15$ and $k=9$. Thus $g=3$, $n'=5$ and $k'=3$.
We have $\ell'=2$ because $2\times 5\equiv1\pmod{3}$.
Thus $\ceil{\tfrac{\ell'n'}{k'}}=4$ and $\floor{\tfrac{\ell'n'}{k'}}=3$.
We have $L_0=0$, $L_1=4$ and $L_2=7$.
A rotation fixes the first $g=3$ onsets (the darker or blue dots).
As shown by the arrows, these onsets imply the positions of the
next three onsets (medium or green dots),
which in turn imply the positions of the final three onsets (the light or yellow dots).}
\end{figure}

To conclude that $R$ is uniquely determined, we must show
that over the range $i\in[0,g-1]$ and $j\in[0,k'-1]$,
the numbers $i k'+j\ell'$ are distinct modulo $k$, and
the numbers $i n'+L_j$ are distinct modulo $n$.

First we show that the numbers $i k'+j\ell'$ are distinct modulo $k$.
Suppose that
\begin{equation}
\label{eqn:DistinctA1}
ik'+j\ell'\equiv pk'+j\ell'\pmod{k}
\end{equation}
for some $i,p\in[0,g-1]$ and $j,q\in[0,k'-1]$.
Because $k=k'\cdot g$, we can write $(ik'+j\ell')\bmod{k}$ as a multiple
of $k'$ plus a residue modulo $k'$. In particular,
\begin{equation*}
(ik'+j\ell')\bmod{k}=
k'\bracket{(i+\floor{\tfrac{j\ell'}{k'}})\bmod{g}}+
(j\ell'\bmod{k'})\enspace.
\end{equation*}
Thus Equation \eqref{eqn:DistinctA1} implies that
\begin{equation}
\label{eqn:DistinctA2}
k'\bracket{(i+\floor{\tfrac{j\ell'}{k'}})\bmod{g}}+(j\ell'\bmod{k'})=
k'\bracket{(p+\floor{\tfrac{q\ell'}{k'}})\bmod{g}}+(q\ell'\bmod{k'})
\enspace.
\end{equation}
Hence $j\ell'\equiv q\ell'\pmod{k'}$.
Thus $j=q$ by Lemma~\ref{lem:Number}(c).
By substituting $j=q$ into Equation~\eqref{eqn:DistinctA2},
it follows that $i\equiv p\pmod{g}$.
Thus $i=p$ because $i,p\in[0,g-1]$.
This proves that the numbers $i k'+j\ell'$ are distinct modulo $k$.

Now we show that the numbers $i n'+L_j$ are distinct modulo $n$.
The proof is similar to the above proof that the numbers $i k'+j\ell'$
are distinct modulo~$k$.

Suppose that
\begin{equation}
\label{eqn:DistinctB1}
in'+L_j\equiv pn'+L_q\pmod{n}
\end{equation}
for some $i,p\in[0,g-1]$ and $j,q\in[0,k'-1]$.
Because $n=n'\cdot g$, we can write $(in'+L_j)\bmod{n}$ as a multiple of $n'$
plus a residue modulo $n'$. In particular,
\begin{equation*}
(in'+L_j)\bmod{n}=
n'\bracket{(i+\FLOOR{\tfrac{L_j}{n'}})\bmod{g}}+
(L_j\bmod{n'})\enspace.
\end{equation*}
Thus Equation \eqref{eqn:DistinctB1} implies that
\begin{equation}
\label{eqn:DistinctB2}
n'\bracket{(i+\FLOOR{\tfrac{L_j}{n'}})\bmod{g}}+
(L_j\bmod{n'})=
n'\bracket{(p+\FLOOR{\tfrac{L_q}{n'}})\bmod{g}}+
(L_q\bmod{n'})\enspace.
\end{equation}
Hence $L_j\equiv L_q\pmod{n'}$. We claim that $j=q$.
If $j=0$ then $L_j=0$, implying $L_q=0$ and $q=0$. Now assume that $j,q\geq1$.
In this case, $L_j=j\floor{\tfrac{\ell'n'}{k'}}+1$ and
$L_q=q\floor{\tfrac{\ell'n'}{k'}}+1$.
Thus
\begin{equation*}
j\FLOOR{\tfrac{\ell'n'}{k'}}\equiv
q\FLOOR{\tfrac{\ell'n'}{k'}}\pmod{k'}\enspace.
\end{equation*}
Hence $j=q$ by Lemma~\ref{lem:Number}(c).
By substituting $j=q$ into Equation~\eqref{eqn:DistinctB2},
it follows that $i\equiv p\pmod{g}$.
Thus $i=p$ because $i,p\in[0,g-1]$.
This proves that the numbers $i n'+L_j$ are distinct modulo $n$.

Therefore $R$ is uniquely determined.
\end{proof}

We have shown that each of the three algorithms generates
a rhythm with property \STAR,
and that there is a unique rhythm with property \STAR.
Thus all of the algorithms produce the same rhythm, up to rotation.
It remains to prove that this rhythm has maximum evenness.

\subsection{Rhythms with Maximum Evenness}
\label{Max_Evenness}


We start with a technical lemma. 
Let $v,w$ be points at geodesic distance $d$ on a circle $C$. 
Obviously $\chordd(v,w)$ is a function of $d$, independent of $v$ and $w$. 
Let $f(C,d) = \chordd(v,w)$.

\begin{lemma} 
\label{lem:trig}
For all geodesic lengths $x\leq d$ on a circle $C$, 
we have $f(C,x)+f(C,d-x)\leq 2\cdot f(C,\frac{d}{2})$, with equality only if $d=2x$.
\end{lemma}

\begin{proof} We can assume that $C$ is a unit circle. 
Consider the isosceles triangle formed by the center of $C$ and a geodesic of length
$d$ ($\leq\pi$). We have $\half f(C,d)=\sin\frac{d}{2}$. 
Thus $f(C,d)=2\sin\frac{d}{2}$. Thus our claim is equivalent to
$\sin x+\sin(d-x)\leq 2\sin\frac{d}{2}$ for all $x\leq d$ ($\leq\pi/2$). 
In the range $0\leq x\leq d$, $\sin x$ is increasing, and
$\sin(d-x)$ is decreasing at the opposite rate. Thus 
$\sin x+\sin(d-x)$ is maximized when $x=d-x$. That is, when $d=2x$. The result follows.
\end{proof}


For a rhythm $R=\{r_0,r_1,\dots,r_{k-1}\}_n$, for each $\ell\in[1,k]$, 
let $S(R,\ell)$ be the sum of chordal distances taken over all ordered 
pairs $(r_i,r_{i+\ell})$ in $R$. 
That is, let $S(R,\ell) =\sum_{i=0}^{k-1}\chordd(r_i,r_{i+\ell})$. 
Property \AAA\ says that $R$ maximizes $\sum_{\ell=1}^kS(R,\ell)$. 
Before we characterize rhythms that maximize the sum of $S(R,\ell)$, 
we first concentrate on rhythms that 
maximize $S(R,\ell)$ for each particular value of $\ell$. 
Let $D(R,\ell)$ be the multiset of clockwise distances $\{\clockwised(r_i,r_{i+\ell}):i\in[0,k-1]\}$. 
Then $S(R,\ell)$ is determined by $D(R,\ell)$. 
In particular, $S(R,\ell)=\sum\{f(C_n,d):d\in D(R,\ell)\}$ (where
$\{f(C_n,d):d\in D(R,\ell)\}$ is a multiset). 

\begin{lemma}
\label{Level}
Let $1\leq\ell\leq k\leq n$ be integers. 
A $k$-onset rhythm $R=\{r_0,r_1,\dots,r_{k-1}\}_n$ maximizes $S(R,\ell)$ 
if and only if $|\clockwised(r_i,r_{i+\ell})-\clockwised(r_j,r_{j+\ell})|\leq1$ 
for all $i,j\in[0,k-1]$.
\end{lemma}

\begin{proof}
Suppose that $R=\{r_0,r_1,\dots,r_{k-1}\}_n$ maximizes $S(R,\ell)$.
Let $d_i = \clockwised(r_i,r_{i+\ell})$ for all $i\in[0,k-1]$. 
Suppose on the contrary that $d_p\geq d_q+2$ for some $p,q\in[0,k-1]$. 
We can assume that $q<p$, $d_p=d_q+2$, and $d_i=d_q+1$ for all $i\in[q+1,p-1]$. 
Define $r'_i = r_i+1$ for all $i\in[q+1,p]$, and define
$r'_i = r_i$ for all other~$i$. 
Let $R'$ be the rhythm $\{r'_0,r'_1,\dots,r'_{k-1}\}_n$. 
Thus $D(R,\ell)\setminus D(R',\ell)=\{d_p,d_q\}$ and 
$D(R',\ell)\setminus D(R,\ell)=\{d_p-1,d_q+1\}$. 
Now $d_p-1=d_q+1=\half(d_p+d_q)$. 
By Lemma~\ref{lem:trig}, $f(C_n,d_p)+f(C_n,d_q)<2\cdot f(C_n,\half(d_p+d_q)$. 
Thus $S(R,\ell)<S(R',\ell)$, which contradicts the maximality of $S(R,\ell)$.

For the converse, let $R$ be a rhythm such that 
$|\clockwised(r_i,r_{i+\ell})-\clockwised(r_j,r_{j+\ell})|\leq1$ for all $i,j\in[0,k-1]$.
Suppose on the contrary that $R$ does not maximize $S(R,\ell)$. Thus
some rhythm $T=(t_0,t_1,\dots,t_{k-1})$ maximizes $S(T,\ell)$ and $T\ne R$. 
Hence $D(T,\ell)\ne D(R,\ell)$.
Because $\sum D(R,\ell)=\sum D(T,\ell)$ ($=\ell n$), 
we have $\clockwised(t_i,t_{i+\ell})-\clockwised(t_j,t_{j+\ell})\geq2$ 
for some $i,j\in[0,k-1]$. 
As we have already proved, this implies that $T$ does not maximize $S(T,\ell)$. 
This contradiction proves that $R$ maximizes $S(R,\ell)$.
\end{proof}

Because $\sum_{i=0}^{k-1}\clockwised(r_i,r_{i+\ell})=\ell n$ 
for any rhythm with $k$ onsets and timespan~$n$, 
Lemma~\ref{Level} can be restated as follows.

\begin{corollary}
\label{LevelLevel}
Let $1\leq\ell\leq k\leq n$ be integers. 
A $k$-onset rhythm $R=\{r_0,r_1,\dots,r_{k-1}\}_n$   
maximizes $S(R,\ell)$ if and only if 
$\clockwised(r_i,r_{i+\ell})\in\{\ceil{\frac{\ell n}{k}},\floor{\frac{\ell n}{k}}\}$ 
for all $i\in[0,k-1]$.\qed
\end{corollary}

\begin{proof}[Proof \STAR\ $\Rightarrow$ \AAA]
If \STAR\ holds for some rhythm $R$, then by 
Corollary~\ref{LevelLevel}, $R$ maximizes $S(R,\ell)$ for \emph{every} $\ell$. 
Thus $R$ maximizes $\sum_\ell S(R,\ell)$.  
\end{proof}

\begin{proof}[Proof \AAA\ $\Rightarrow$ \STAR]
By Theorem~\ref{thm:Uniqueness}, there is a unique rhythm $R$ that 
satisfies property \STAR.
Let $R$ denote the unique rhythm that satisfies property \STAR. 
Suppose on the contrary that there is a rhythm $T=(t_0,t_1,\dots,t_{k-1})$ 
with property \AAA\ but $R\neq T$. Thus there exists an ordered pair 
$(t_i,t_{i+\ell})$ in $T$ with clockwise distance 
$\clockwised(t_i,t_{i+\ell})\not\in\{\floor{\frac{\ell n}{k}},\ceil{\frac{\ell n}{k}}\}$. 
By Corollary~\ref{LevelLevel}, $S(T,\ell)<S(R,\ell)$. 
Because $T$ has property \AAA, $\sum_{\ell=1}^k S(T,\ell)\geq\sum_{\ell=1}^k S(R,\ell)$. 
Thus for some $\ell'$ we have $S(T,\ell')>S(R,\ell')$. 
But this is a contradiction, because $S(R,\ell')\geq S(T,\ell')$ by Corollary~\ref{LevelLevel}.
\end{proof}

This completes the proof of Theorem~\ref{Main_Even}.
Note that Theorem~\ref{Main_Even} holds for metrics other
than the pairwise sum of all chordal distances. The only
property that is needed is Lemma~\ref{lem:trig}. For example,
the metric ``pairwise sum of the squares of all geodesic distances''
satisfies Lemma~\ref{lem:trig}, and thus Theorem~\ref{Main_Even} holds
in this setting.


\section{Deep Rhythms}
\label{sec:Deep}

Winograd \cite{Winograd-1966}, and independently
Clough et al.~\cite{clough-99b},
characterize all Winograd-deep scales:
up to rotation, they are the scales that can be generated by
the first $\lfloor n/2 \rfloor$ or $\lfloor n/2 \rfloor+1$ multiples
(modulo~$n$) of a value that is relatively prime
to~$n$, plus one exceptional scale $\{0,1,2,4\}_6$.
In this section, we prove a similar (but more general) characterization of
Erd\H{o}s-deep rhythms:
up to rotation and scaling, they are the rhythms generable
as the first $k$ multiples (modulo~$n$) of a value that is relatively
prime to~$n$, plus the same exceptional rhythm $\{0,1,2,4\}_6$.
The key difference is that the number of onsets $k$ is now a free parameter,
instead of being forced to be either $\lfloor n/2 \rfloor$ or
$\lfloor n/2 \rfloor + 1$.
Our proof follows Winograd's, but differs
in one case (the second case of Theorem~\ref{deep characterization}).

We later prove that every Erd\H{o}s-deep rhythm has a shelling and
that maximally even rhythms with $n$ and $k$ relatively prime
are Erd\H{o}s-deep.

\subsection{Characterization of Deep Rhythms}

Our characterization of Erd\H{o}s-deep rhythms is
in terms of two families of rhythms.
The main rhythm family consists of the generated rhythms
$D_{k,n,m} = \{i m \bmod n : i = 0, 1, \dots, k-1\}_n$
of timespan~$n$, for certain values of $k$, $n$, and~$m$.
The one exceptional rhythm is $F = \{0,1,2,4\}_6$ of timespan~$6$.

\begin{fact} \label{F is deep}
  $F$ is Erd\H{o}s-deep.
\end{fact}

\begin{lemma} \label{D is deep}
  If $k \leq \lfloor n/2 \rfloor + 1$ and $m$ and $n$ are relatively prime, 
  then $D_{k,n,m}$ is Erd\H{o}s-deep.
  %
\end{lemma}

\begin{proof}
  The multiset of clockwise distances in $D_{k,n,m}$ is
  $\{(j m - i m) \bmod n : i < j\} =
  \{(j - i) m \bmod n : i < j\}$.
  There are $k-p$ choices of $i$ and $j$ such that $j-i = p$,
  so there are exactly $p$ occurrences of the clockwise distance
  $(p m) \bmod n$ in the multiset.
  Each of these clockwise distances corresponds to a geodesic
  distance---either $(p m) \bmod n$ or $(- p m) \bmod n$,
  whichever is smaller (at most~$n/2$).
  We claim that these geodesic distances are all distinct.
  Then the multiplicity of each geodesic distance $(\pm p m) \bmod n$
  is exactly $p$, establishing that the rhythm is Erd\H{o}s-deep.
  
  For two geodesic distances to be equal, we must have
  $\pm p m \equiv \pm q m \pmod{n}$ for some (possibly different)
  choices for the $\pm$ symbols, and for some $p \neq q$.
  By (possibly) multiplying both sides by $-1$, we obtain two cases:
  (1)~$p m \equiv  q m \pmod{n}$ and
  (2)~$p m \equiv -q m \pmod{n}$.
  Because $m$ is relatively prime to~$n$, by Lemma~\ref{lem:Inverse},
  $m$~has a multiplicative inverse modulo~$n$.
  Multiplying both sides of the congruence by this inverse, we obtain
  (1)~$p \equiv  q \pmod{n}$ and
  (2)~$p \equiv -q \pmod{n}$.
  Because $0 \leq i < j < k \leq \lfloor n/2 \rfloor + 1$, we have
  $0 \leq p = j-i < \lfloor n/2 \rfloor + 1$, and similarly for~$q$:
  $0 \leq p, q \leq \lfloor n/2 \rfloor$.
  Thus, the first case of $p \equiv q \pmod{n}$ can happen only when $p = q$,
  and the second case of $p+q \equiv 0 \pmod{n}$ can happen only when
  $p = q = 0$ or when $p = q = n/2$.
  Either case contradicts that $p \neq q$.
  Therefore the geodesic distances arising from different values of $p$
  are indeed distinct, proving the lemma.
\end{proof}

We now state and prove our characterization of Erd\H{o}s-deep rhythms,
which is up to rotation and scaling.
Rotation preserves the geodesic distance multiset and therefore
Erd\H{o}s-deepness (and Winograd-deepness).
Scaling maps each geodesic distance $d$ to $\alpha d$, and thus preserves
multiplicities and therefore Erd\H{o}s-deepness (but not Winograd-deepness).

\begin{theorem} \label{deep characterization}
  A rhythm is Erd\H{o}s-deep if and only if it is a rotation of a scaling of
  either the rhythm $F$ or the rhythm $D_{k,n,m}$ for some $k,n,m$ with
  $k \leq \lfloor n/2 \rfloor + 1$, $1 \leq m \leq \lfloor n/2 \rfloor$,
  and $m$ and $n$ are relatively prime.
\end{theorem}

\begin{proof}
  Because a rotation of a scaling of an Erd\H{o}s-deep rhythm is
  Erd\H{o}s-deep,
  the ``if'' direction of the theorem follows from
  Fact \ref{F is deep} and Lemma \ref{D is deep}.

  Consider an Erd\H{o}s-deep rhythm~$R$ with $k$ onsets.
  By the definition of Erd\H{o}s-deepness,
  $R$ has one nonzero geodesic distance with multiplicity $i$
  for each $i = 1, 2, \dots, k-1$.
  Let $m$ be the geodesic distance with multiplicity $k-1$.
  Because $m$ is a geodesic distance, $1 \leq m \leq \lfloor n/2 \rfloor$.
  Also, $k \leq \lfloor n/2 \rfloor + 1$ (for any Erd\H{o}s-deep rhythm~$R$),
  because all nonzero geodesic distances are between $1$ and $\lfloor n/2 \rfloor$
  and therefore at most $\lfloor n/2 \rfloor$ nonzero geodesic distances occur.
  Thus $k$ and $m$ are suitable parameter choices for $D_{k,n,m}$.

  Consider the graph $G_m = (R, E_m)$ with vertices corresponding to onsets
  in $R$ and with an edge between two onsets of geodesic distance~$m$.
  By the definition of geodesic distance, every vertex $i$ in $G_m$ has degree at most~$2$:
  the only onsets at geodesic distance exactly $m$ from $i$ are $(i-m) \bmod n$ and
  $(i+m) \bmod n$.
  Thus, the graph $G_m$ is a disjoint union of paths and cycles.
  The number of edges in $G_m$ is the multiplicity of $m$, which we supposed
  was $k-1$, which is $1$ less than the number of vertices in~$G_m$.
  Thus, the graph $G_m$ consists of exactly one path and any number of cycles.

  The cycles of $G_m$ have a special structure because they correspond
  to subgroups generated by single elements in the cyclic group
  $(\mathbb{Z}/(n), +)$.
  Namely, the onsets corresponding to vertices of a cycle in $G_m$
  form a regular $(n/a)$-gon, with a geodesic distance of $a = \gcd(m,n)$
  between consecutive onsets.
  ($a$ is called the index of the subgroup generated by~$m$.)
  In particular, every cycle in $G_m$ has the same length $r = n/a$.
  Because $G_m$ is a simple graph, every cycle must have at least $3$
  vertices, so $r \geq 3$.

  The proof partitions into four cases depending on the length of the path
  and on how many cycles the graph $G_m$ has.
  The first two cases will turn out to be impossible;
  the third case will lead to a rotation of a scaling of rhythm~$F$; and
  the fourth case will lead to a rotation of a scaling of rhythm~$D_{k,n,m}$.

  First suppose that the graph $G_m$ consists of a path of length at least $1$
  and at least one cycle.
  We show that this case is impossible
  because the rhythm $R$ can have no geodesic distance with multiplicity~$1$.
  Suppose that there is a geodesic distance with multiplicity~$1$,
  say between onsets $i_1$ and~$i_2$.
  If $i$ is a vertex of a cycle,
  then both $(i+m) \bmod n$ and $(i-m) \bmod n$ are onsets in~$R$.
  If $i$ is a vertex of the path, then one or two of these are onsets in~$R$,
  with the case of one occurring only at the endpoints of the path.
  If $(i_1+m) \bmod n$ and $(i_2+m) \bmod n$ were both onsets in~$R$,
  or $(i_1-m) \bmod n$ and $(i_2-m) \bmod n$ were both onsets in~$R$, then
  we would have another occurrence of the geodesic distance between $i_1$ and~$i_2$,
  contradicting that this geodesic distance has multiplicity~$1$.
  Thus, $i_1$ and $i_2$ must be opposite endpoints of the path.
  If the path has length $\ell$, then the clockwise distance
  between $i_1$ and $i_2$ is $(\ell m) \bmod n$.
  This clockwise distance (and hence the corresponding geodesic distance)
  appears in every cycle, of which there is at least one,
  so the geodesic distance has multiplicity more than~$1$, a contradiction.
  Therefore this case is impossible.

  Second suppose that the graph $G_m$ consists of a path of length $0$
  and at least two cycles.
  We show that this case is impossible
  because the rhythm $R$ has two geodesic distances with the same multiplicity.
  Pick any two cycles $C$ and~$C'$, and let $d$ be the smallest positive
  clockwise distance from a vertex of $C$ to a vertex of~$C'$.
  Thus $i$ is a vertex of $C$ if and only if $(i+d) \bmod n$
  is a vertex of~$C'$.
  Because the cycles are disjoint, $d < a$.
  Because $r \geq 3$, $d < n/3$, so clockwise distances of $d$ are also 
  geodesic distances of~$d$.
  The number of occurrences of  geodesic distance $d$ between a vertex of $C$
  and a vertex of $C'$ is either $r$ or~$2r$, the case of $2r$ arising
  when $d = a/2$ (that is, $C'$ is a ``half-rotation'' of~$C$).
  The number of occurrences of geodesic distance $d' = \min\{d+m, n-(d+m)\}$
  is the same---either $r$ or $2r$, in the same cases.
  (Note that $d < a \leq n-m$, so $d+m < n$, so the definition of $d'$
  correctly captures a geodesic distance modulo~$n$.)
  The same is true of geodesic distance $d'' = \min\{d-m, n-(d-m)\}$.
  If other pairs of cycles have the same smallest positive clockwise
  distance~$d$, then the number of occurrences of $d$, $d'$, and $d''$
  between those cycles are also equal.
  Because the cycles are disjoint, geodesic distance $d$ and thus $d+m$ and $d-m$
  cannot be $(p m) \bmod n$ for any~$p$,
  so these geodesic distances cannot occur between two vertices of the same cycle.
  Finally, the sole vertex $x$ of the path has geodesic distance $d$ to onset $i$
  (which must be a vertex of some cycle) if and only if $x$ has geodesic distance
  $d'$ to onset $(i+m) \bmod n$ (which must be a vertex of the same cycle)
  if and only if $x$ has geodesic distance $d''$ to onset $(i-m) \bmod n$
  (which also must be a vertex of the same cycle).
  Therefore the multiplicities of geodesic distances $d$, $d'$, and $d''$ must be equal.
  Because $R$ is Erd\H{o}s-deep, we must have $d = d' = d''$.
  To have $d = d'$, either $d = d+m$ or $d = n-(d+m)$,
  but the first case is impossible because $d > 0$
  by nonoverlap of cycles, so $2 d + m = n$.
  Similarly, to have $d = d''$, we must have $2 d - m = n$.
  Subtracting these two equations, we obtain that $2 m = 0$,
  contradicting that $m > 0$.
  Therefore this case is also impossible.

  Third suppose that the graph $G_m$ consists of a path of length $0$
  and exactly one cycle.
  We show that this case forces $R$ to be a rotation of a scaling of rhythm $F$
  because otherwise two geodesic distances $m$ and $m'$ have the same multiplicity.
  The number of occurrences of geodesic distance $m$ in the cycle
  is precisely the length $r$ of the cycle.
  Similarly, the number of occurrences of geodesic distance
  $m' = \min\{2 m, n - 2 m\}$ in the cycle is~$r$.
  The sole vertex $x$ on the path cannot have geodesic distance $m$ or $m'$
  to any other onset (a vertex of the cycle)
  because then $x$ would then be on the cycle.
  Therefore the multiplicities of geodesic distances $m$ and $m'$ must be equal.
  Because $R$ is Erd\H{o}s-deep, $m$ must equal~$m'$, which implies that
  either $m = 2 m$ or $m = n - 2 m$.
  The first case is impossible because $m > 0$.
  In the second case, $3 m = n$, that is, $m = {1 \over 3} n$.
  Therefore, the cycle has $r=3$ vertices, say at
  $\Delta, \Delta + {1 \over 3} n, \Delta + {2 \over 3} n$.
  The fourth and final onset $x$ must be midway between two of these three
  onsets, because otherwise its geodesic distance to the three vertices are all
  distinct and therefore unique.
  No matter where $x$ is so placed, the rhythm $R$ is a rotation
  by $\Delta + c {1 \over 3} n$ (for some $c \in \{0,1,2\}$)
  of a scaling by $n/6$ of the rhythm~$F$.

  Finally suppose that $G_m$ has no cycles, and consists solely of a path.
  We show that this case forces $R$ to be a rotation of a scaling of
  a rhythm $D_{k,n',m'}$ with $1 \leq m' \leq \lfloor n'/2 \rfloor$
  and with $m'$ and $n'$ relatively prime.
  Let $b$ be the onset such that $(b-m) \bmod n$ is not an onset
  (the ``beginning'' vertex of the path).
  Consider rotating $R$ by $-i$ so that $0$ is an onset in the resulting
  rhythm~$R-i$.
  The vertices of the path in $R-i$ form a subset of the subgroup of the
  cyclic group $(\mathbb{Z}/(n),+)$ generated by the element~$m$.  
  Therefore the rhythm
  $R-i = D_{k,n,m} = \{(i m) \bmod n : i = 0, 1, \dots, k-1\}_n$
  is a scaling by $a$ of the rhythm
  $D_{k,n/a,m/a} = \{(i m/a) \bmod (n/a) : i = 0, 1, \dots, k-1\}_n$.
  The rhythm $D_{k,n/a,m/a}$ has an appropriate value for the third argument:
  $m/a$ and $n/a$ are relatively prime ($a = \gcd(m,n)$) and
  $1 \leq m/a \leq \lfloor n/2 \rfloor / a \leq \lfloor (n/a)/2 \rfloor$.
  Also, $k \leq \lfloor (n/a)/2 \rfloor + 1$ because the only occurring
  geodesic distances are multiples of $a$ and therefore the number $k-1$ of distinct
  geodesic distances is at most $\lfloor (n/a)/2 \rfloor$.
  Therefore $R$ is a rotation by $i$ of a scaling by $a$ of $D_{k,n/a,m/a}$
  with appropriate values of the arguments.
\end{proof}

\begin{corollary}
  A rhythm is Erd\H{o}s-deep if and only if it is a rotation of a scaling of
  the rhythm $F$ or it is a rotation of a rhythm $D_{k,n,m}$ for some $k,n,m$
  satisfying $k \leq \lfloor n/2 g \rfloor + 1$ where $g = \gcd(m,n)$.
\end{corollary}

\begin{proof}
  First we show that any Erd\H{o}s-deep rhythm has one of the two forms
  in the corollary.  By Theorem~\ref{deep characterization}, there are
  two flavors of Erd\H{o}s-deep rhythms, and the corollary directly handles
  rotations of scalings of~$F$.  Thus it suffices to consider
  a rhythm $R$ that is a rotation by $\Delta$ of a scaling by $\alpha$ of
  $D_{k,n,m}$ where $k \leq \lfloor n/2 \rfloor + 1$,
  $1 \leq m \leq \lfloor n/2 \rfloor$,
  and $m$ and $n$ are relatively prime.
  Equivalently, $R$ is a rotation by $\Delta$ of $D_{k,n',m'}$
  where $n' = \alpha n$ and $m' = \alpha m$.
  Now $g = \gcd(n',m') = \alpha$, so $n'/ g = n$.
  Hence, $k \leq \lfloor n'/2 g \rfloor + 1$ as desired.
  Thus we have rewritten $R$ in the desired form.

  It remains to show that every rhythm in one of the two forms
  in the corollary is Erd\H{o}s-deep.  Again, rotations of scalings of $F$
  are handled directly by Theorem~\ref{deep characterization}.
  So consider a rotation of $D_{k,n,m}$
  where $k \leq \lfloor n/2g \rfloor + 1$.
  The value of $m$ matters only modulo~$n$,
  so we assume that $0 \leq m \leq n-1$.

  First we show that, if $\lfloor n/2 \rfloor + 1 \leq m \leq n-1$,
  then $D_{k,n,m}$ can be rewritten as a rotation of the rhythm $D_{k,n,m'}$
  where $m' = n - m \leq \lfloor n/2 \rfloor$. 
  By reversing the order in which we list the onsets in
  $D_{k,n,m} = \{i m \bmod n : i = 0, 1, \dots, k-1\}_n$,
  we can write
  $D_{k,n,m} = \{ (k - 1 - i) \, m \bmod n : i = 0, 1, \dots, k-1\}_n$.
  Now consider rotating the rhythm
  $D_{k,n,n-m} = \{ i \, (n - m) \bmod n : i = 0, 1, \dots, k-1\}_n$
  by $(k-1) m$.
  We obtain the rhythm
  $\{[i \, (n - m) + (k-1) \, m] \bmod n : i = 0, 1, \dots, k-1\}_n =
   \{[(k - 1 - i) \, m + i n] \bmod n : i = 0, 1, \dots, k-1\}_n =
   \{(k - 1 - i) \, m \bmod n : i = 0, 1, \dots, k-1\}_n = D_{k,n,m}$
  as desired.

  Thus it suffices to consider rotations of $D_{k,n,m}$
  where $1 \leq m \leq \lfloor n/2 \rfloor$ and $k \leq \lfloor n/2 g \rfloor$.
  The rhythm $D_{k,n',m'}$, where $n' = n/g$ and $m' = m/g$,
  is Erd\H{o}s-deep by Theorem~\ref{deep characterization}
  because $n'$ and $m'$ are relatively prime,
  $k \leq \lfloor n'/2 \rfloor + 1$, and
  $1 \leq m' \leq \lfloor n'/2 \rfloor$.
  But $D_{k,n,m}$ is the scaling of $D_{k,n',m'}$ by the integer~$g$,
  so $D_{k,n,m}$ is also Erd\H{o}s-deep.
\end{proof}

An interesting consequence of this characterization is the following:

\begin{corollary}
  Every Erd\H{o}s-deep rhythm has a shelling.
\end{corollary}

\begin{proof}
  If the Erd\H{o}s-deep rhythm is $D_{k,n,m}$, we can remove the last onset
  from the path, resulting in $D_{k-1,n,m}$, and repeat until we obtain the
  empty rhythm $D_{0,n,m}$.
  At all times, $k$ remains at most $\lfloor n/2 \rfloor + 1$
  (assuming it was originally) and $m$ remains between $1$ and
  $\lfloor n/2 \rfloor$ and relatively prime to~$n$.
  On the other hand, $F = \{0,1,2,4\}_6$ has the shelling
  $4, 2, 1, 0$ because $\{0,1,2\}_6$ is Erd\H{o}s-deep.
\end{proof}



\subsection{Connection Between Deep and Even Rhythms}

A connection between maximally even scales and Winograd-deep scales is shown 
by Clough et al.~\cite{clough-99b}. 
They define a \emph{diatonic scale} to be a maximally even scale with 
$k = (n+2)/2$ and $n$ a multiple of~$4$. 
They show that diatonic scales are Winograd-deep. 
We now prove a similar result for Erd\H{o}s-deep rhythms.

\begin{lemma}
\label{Erd deep}
  A rhythm $R$ of maximum evenness satsifying $k \leq  \lfloor n/2 \rfloor + 1$
  is Erd\H{o}s-deep if and only if $k$ and $n$ are relatively prime.
\end{lemma}

\begin{proof}
Recall that by property \STAR\, one of the unique characterizations of an
even rhythm of maximum evenness can be stated as follows.
For all $1 \leq \ell \leq k$, and for every ordered pair $(r_i,r_{i+\ell})$ 
of onsets in~$R$, the clockwise distance 
$\clockwised(r_i,r_{i+\ell})\in\{\lfloor {\frac{\ell n}{k}} \rfloor, 
\lceil{\frac{\ell n}{k}} \rceil \}$.

For the case in which $k$ and $n$ are relatively prime,
by Lemma~\ref{lem:Inverse},
there exists a value $\ell <  k$ such that $\ell n \equiv 1 \pmod{k}$. 
Thus we can write $ \ell n = k \lfloor  \ell n / k \rfloor + 1$. 
Let $m = \lfloor \ell n /k \rfloor$. 
Now consider the set $\{i m \bmod n : i = 0, 1, \dots, k-1\}_n$. 
By Lemma~\ref{lem:Number}(c), we get $k$ distinct values, so
$R$ can be realized as $D_{k,n,m} = \{i m \bmod n : i = 0, 1, \dots, k-1\}_n$. 
Thus, by Lemma~\ref{D is deep}, $R$ is Erd\H{o}s-deep.

Observe that $F = \{0,1,2,4\}_6$ does not maximize evenness because
$\clockwised(0,2) = 2$ and $\clockwised(2,0) = 4$ yet $\ell=2$.
Hence, any rhythm that maximizes evenness and that is deep
must also be generated. 

Now consider the case in which $n$ and $k$ are not relatively prime. 
We show that the assumption that $R$ is deep leads to a contradiction. 
Thus, assuming that $R$ is deep implies that there is a value $m$  
such that $R$ can be realized as $D_{k,n,m} = \{i m \bmod n : i = 0, 1, \dots, k-1\}_n$. 
This in turn implies that there exists an integer $\ell$ such that $km = \ell n + 1$, 
that is, $\ell n \equiv 1 \pmod{k}$. 
However, for this to happen, $n$ and $k$ must be relatively prime,
a contradiction. 

Thus we have shown that $R$ is Erd\H{o}s-deep if and only if
$k$ and $n$ are relatively prime.
\end{proof}

\label{last technical section}

\section*{Acknowledgments}

This work was initiated at the 20th Bellairs Winter Workshop on
Computational Geometry held January 28--February 4, 2005 
at Bellairs Research Institute in Barbados.
We thank the other participants of that workshop---Greg Aloupis,
David Bremner, Justin Colannino, Mirela Damian, Vida Dujmovi\'c,
Jeff Erickson, Ferran Hurtado, John Iacono, Danny Krizanc, Stefan Langerman,
Erin McLeish, Pat Morin, Mark Overmars, Suneeta Ramaswami, Diane Souvaine,
Ileana Streinu, Remco Veltcamp, and Sue Whitesides---for helpful discussions
and contributing to a fun and creative atmosphere.

Thanks to Simha Arom for providing us with a copy of his paper
on the classification of \emph{aksak} rhythms~\cite{arom-04}, which
provided several examples of Euclidean rhythms of which we were not aware.
Thanks to Jeff Erickson for bringing to our attention the
wonderful paper by Mitchell Harris and Ed Reingold~\cite{harris-04}
on digital line drawing, leap year calculations, and Euclid's algorithm.
Finally, thanks to Marcia Ascher for pointing out
that the leap-year pattern of the Jewish
calendar is a Euclidean rhythmic necklace.

\let\realbibitem=\bibitem
\def\bibitem{\par \vspace{-1.2ex}\realbibitem}


\begin{thebibliography}{DBFG{\etalchar{+}}04}

\bibitem[Aga86]{agawu-86}
V.~Kofi Agawu.
\newblock {G}i {D}unu, {N}yekpadudo, and the study of {W}est {A}frican rhythm.
\newblock {\em Ethnomusicology}, 30(1):64--83, Winter 1986.

\bibitem[Ank97]{anku-97}
Willie Anku.
\newblock Principles of rhythm integration in {A}frican music.
\newblock {\em Black Music Research Journal}, 17(2):211--238, Autumn 1997.

\bibitem[Ape60]{apel-60}
Willi Apel.
\newblock Vier plus vier = drei plus drei plus zwei.
\newblock {\em Acta Musicologica}, 32(Fasc. 1):29--33, January-March 1960.

\bibitem[Aro91]{arom-91}
Simha Arom.
\newblock {\em {A}frican Polyphony and Polyrhythm}.
\newblock Cambridge University Press, Cambridge, England, 1991.

\bibitem[Aro04]{arom-04}
Simha Arom.
\newblock L'aksak: {P}rincipes et typologie.
\newblock {\em Cahiers de Musiques Traditionnelles}, 17:12--48, 2004.

\bibitem[AS02]{allouche-02}
Jean-Paul Allouche and Jeffrey~O. Shallit.
\newblock {\em Automatic Sequences}.
\newblock Cambridge University Press, Cambridge, England, 2002.

\bibitem[Asc75]{asch-75}
Michael~I. Asch.
\newblock Social context and the musical analysis of {S}lavey drum dance songs.
\newblock {\em Ethnomusicology}, 19(2):245--257, May 1975.

\bibitem[Asc02]{ascher-02}
Marcia Ascher.
\newblock {\em Mathematics Elsewhere: An Exploration of Ideas Across Cultures}.
\newblock Princeton University Press, Princeton and Oxford, 2002.

\bibitem[Ash03]{ashton-03}
Anthony Ashton.
\newblock {\em Harmonograph--A Visual Guide to the Mathematics of Music}.
\newblock Walker and Company, New York, 2003.

\bibitem[Bar81]{bartok-81}
B\'ela Bart\'ok.
\newblock Ce qu'on appelle le rythme {b}ulgare.
\newblock In {\em Musique de la vie}, pages 142--155, Stock, Paris, 1981.

\bibitem[Bar04a]{barbour-04}
James~M. Barbour.
\newblock {\em Tuning and Temperament: A Historical Survey}.
\newblock Dover, New York, 2004.

\bibitem[Bar04b]{barz-04}
Gregory Barz.
\newblock {\em Music in East Africa}.
\newblock Oxford University Press, Oxford, England, 2004.

\bibitem[BD94]{block-94}
Steven Block and Jack Douthett.
\newblock Vector products and intervallic weighting.
\newblock {\em Journal of Music Theory}, 38:21--41, 1994.

\bibitem[Beh73]{behague-73}
Gerard Behague.
\newblock Bossa and bossas: recent changes in {B}razilian urban popular music.
\newblock {\em Ethnomusicology}, 17(2):209--233, 1973.

\bibitem[Bek05]{bektas-05}
Tolga Bekta\c{s}.
\newblock Relationships between prosodic and musical meters in the {B}este form
  of classical {T}urkish music.
\newblock {\em Asian Music}, 36(1):1--26, Winter/Spring 2005.

\bibitem[Bjo03a]{bjorklund-03b}
Eric Bjorklund.
\newblock A metric for measuring the evenness of timing system rep-rate
  patterns.
\newblock SNS ASD Technical Note SNS-NOTE-CNTRL-100, Los Alamos National
  Laboratory, Los Alamos, U.S.A., 2003.

\bibitem[Bjo03b]{bjorklund-03a}
Eric Bjorklund.
\newblock The theory of rep-rate pattern generation in the {SNS} timing system.
\newblock SNS ASD Technical Note SNS-NOTE-CNTRL-99, Los Alamos National
  Laboratory, Los Alamos, U.S.A., 2003.

\bibitem[Br{\u{a}}51]{brailoiu-51}
Constantin Br{\u{a}}iloiu.
\newblock Le rythme aksak.
\newblock {\em Revue de Musicologie}, 33:71--108, 1951.

\bibitem[Bra59]{brandel-59}
Rose Brandel.
\newblock The {A}frican hemiola style.
\newblock {\em Ethnomusicology}, 3(3):106--117, September 1959.

\bibitem[Bre65]{bresenham-65}
Jack~E. Bresenham.
\newblock Algorithm for computer control of digital plotter.
\newblock {\em IBM Systems Journal}, 4:25--30, 1965.

\bibitem[Bre99]{brewer-99}
Roy Brewer.
\newblock The use of {H}abanera rhythm in rockabilly music.
\newblock {\em American Music}, 17:300--317, Autumn 1999.

\bibitem[Bru64]{brun-64}
Viggo Brun.
\newblock Euclidean algorithms and musical theory.
\newblock {\em Enseignement Math{\'e}matique}, 10:125--137, 1964.

\bibitem[Car98]{carey-98}
Norman Carey.
\newblock {\em Distribution modulo 1 and musical scales}.
\newblock PhD thesis, University of Rochester, Rochester, New York, 1998.

\bibitem[CD91]{clough-91}
John Clough and Jack Douthett.
\newblock Maximally even sets.
\newblock {\em Journal of Music Theory}, 35:93--173, 1991.

\bibitem[CEK99]{clough-99b}
John Clough, Nora Engebretsen, and Jonathan Kochavi.
\newblock Scales, sets, and interval cycles: a taxonomy.
\newblock {\em Music Theory Spectrum}, 21(1):74--104, Spring 1999.

\bibitem[Che79]{chernoff-79}
John~Miller Chernoff.
\newblock {\em African Rhythm and African Sensibility}.
\newblock The University of Chicago Press, Chicago, 1979.

\bibitem[Che02]{chemillier-02}
Marc Chemillier.
\newblock Ethnomusicology, ethnomathematics. {T}he logic underlying orally
  transmitted artistic practices.
\newblock In Gerard Assayag, Hans~Georg Feichtinger, and Jose~Francisco
  Rodrigues, editors, {\em Mathematics and Music}, pages 161--183.
  Springer-Verlag, 2002.

\bibitem[Cla00]{clayton-00}
Martin Clayton.
\newblock {\em Time in Indian Music}.
\newblock Oxford University Press, New York, 2000.

\bibitem[Cle94]{cler-94}
J\'er\^ome Cler.
\newblock Pour une th\'eorie de l'aksak.
\newblock {\em Revue de Musicologie}, 80:181--210, 1994.

\bibitem[CLRS01]{cormen-01}
Thomas~H. Cormen, Charles~E. Leiserson, Ronald~L. Rivest, and Clifford Stein.
\newblock {\em Introduction to Algorithms}.
\newblock The MIT Press, Cambridge, Massachussetts, 2001.

\bibitem[CM85]{clough-85}
John Clough and Gerald Myerson.
\newblock Variety and multiplicity in diatonic systems.
\newblock {\em Journal of Music Theory}, 29:249--270, 1985.

\bibitem[CM86]{clough-86}
John Clough and Gerald Myerson.
\newblock Musical scales and the generalized circle of fifths.
\newblock {\em American Mathematical Monthly}, 93(9):695--701, 1986.

\bibitem[Cox62]{coxeter-62}
H.~S.~M. Coxeter.
\newblock Music and mathematics.
\newblock {\em The Canadian Music Journal}, VI:13--24, 1962.

\bibitem[CT03]{chemillier-03}
Marc Chemillier and Charlotte Truchet.
\newblock Computation of words satisfying the ``rhythmic oddity property''
  (after {S}imha {A}rom's works).
\newblock {\em Information Processing Letters}, 86:255--261, 2003.

\bibitem[DBFG{\etalchar{+}}04]{banez-04}
Miguel D{\'i}az-Ba{\~n}ez, Giovanna Farigu, Francisco G{\'o}mez, David
  Rappaport, and Godfried~T. Toussaint.
\newblock El comp{\'a}s flamenco: a phylogenetic analysis.
\newblock In {\em Proceedings of BRIDGES: Mathematical Connections in Art,
  Music and Science}, Southwestern College, Winfield, Kansas, July 30 - August
  1 2004.

\bibitem[EMF90]{el-mallah-90}
Issam El-Mallah and Kai Fikentscher.
\newblock Some observations on the naming of musical instruments and on the
  rhythm in {O}man.
\newblock {\em Yearbook for Traditional Music}, 22:123--126, 1990.

\bibitem[Erd89]{Erdos-1989}
Paul Erd{\H{o}}s.
\newblock Distances with specified multiplicities.
\newblock {\em American Mathematical Monthly}, 96:447, 1989.

\bibitem[ERSS03]{ellis-03}
John Ellis, Frank Ruskey, Joe Sawada, and Jamie Simpson.
\newblock Euclidean strings.
\newblock {\em Theoretical Computer Science}, 301:321--340, 2003.

\bibitem[Euc56]{euclid-56}
Euclid.
\newblock {\em Elements}.
\newblock Dover, 1956.
\newblock Translated by Sir Thomas L.~Heath.

\bibitem[Eva66]{evans-66}
Bob Evans.
\newblock {\em Authentic Conga Rhythms}.
\newblock Belwin Mills Publishing Corporation, Miami, 1966.

\bibitem[Far92]{farquharson-92}
Mary Farquharson.
\newblock {\em Africa in America}.
\newblock Discos Corazon, Mexico, 1992.
\newblock [CD].

\bibitem[Flo99]{floyd-99}
Samuel~A. Floyd, Jr.
\newblock Black music in the circum-{C}aribbean.
\newblock {\em American Music}, 17(1):1--38, 1999.

\bibitem[For73]{forte-73}
Allen Forte.
\newblock {\em The Structure of Atonal Music}.
\newblock Yale Univ. Press, New Haven, Connecticut, U.S.A., 1973.

\bibitem[Fra56]{franklin-56}
Philip Franklin.
\newblock The {E}uclidean algorithm.
\newblock {\em The American Mathematical Monthly}, 63(9):663--664, November
  1956.

\bibitem[Gam67a]{Gamer-1967b}
Carlton Gamer.
\newblock Deep scales and difference sets in equal-tempered systems.
\newblock In {\em Proceedings of the 2nd Annual Conference of the American
  Society of University Composers}, pages 113--122, 1967.

\bibitem[Gam67b]{Gamer-1967}
Carlton Gamer.
\newblock Some combinational resources of equal-tempered systems.
\newblock {\em Journal of Music Theory}, 11:32--59, 1967.

\bibitem[Gam02]{gamboa-02}
Jos{\'e}~Manuel Gamboa.
\newblock {\em Cante por Cante: Discolibro Didactico de Flamenco}.
\newblock New Atlantis Music, Alia Discos, Madrid, 2002.

\bibitem[GMRT05]{gomez-05}
Francisco G{\'o}mez, Andrew Melvin, David Rappaport, and Godfried~T. Toussaint.
\newblock Mathematical measures of syncopation.
\newblock In {\em Proc. BRIDGES: Mathematical Connections in Art, Music and
  Science}, pages 73--84, Banff, Alberta, Canada, July 31 - August 3 2005.

\bibitem[HAD95]{hartigan-95}
Royal Hartigan, Abraham Adzenyah, and Freeman Donkor.
\newblock {\em West African Rhythms for Drum Set}.
\newblock Manhattan Music, 1995.

\bibitem[Hag03]{hagoel-03}
Kobi Hagoel.
\newblock {\em The Art of Middle Eastern Rhythm}.
\newblock OR-TAV Music Publications, Kfar Sava, Israel, 2003.

\bibitem[Has97]{hasty-97}
Christopher~F. Hasty.
\newblock {\em Meter as Rhythm}.
\newblock Oxford University Press, Oxford, England, 1997.

\bibitem[HK81]{hye-ku-81}
Lee Hye-Ku.
\newblock Quintuple meter in {K}orean instrumental music.
\newblock {\em Asian Music}, 13(1):119--129, 1981.

\bibitem[HR04]{harris-04}
Mitchell~A. Harris and Edward~M. Reingold.
\newblock Line drawing, leap years, and {E}uclid.
\newblock {\em ACM Computing Surveys}, 36(1):68--80, March 2004.

\bibitem[Joh03]{Johnson-2003}
Timothy~A. Johnson.
\newblock {\em Foundations of Diatonic Theory: A Mathematically Based Approach
  to Music Fundamentals}.
\newblock Key College Publishing, Emeryville, California, 2003.

\bibitem[Kau80]{kauffman-80}
Robert Kauffman.
\newblock {A}frican rhythm: A reassessment.
\newblock {\em Ethnomusicology}, 24(3):393--415, September 1980.

\bibitem[Kei91]{keith-91}
Michael Keith.
\newblock {\em From Polychords to P{\' o}lya: Adventures in Musical
  Combinatorics}.
\newblock Vinculum Press, Princeton, 1991.

\bibitem[Kl{\H o}97]{klower-97}
T{\H o}m Kl{\H o}wer.
\newblock {\em The Joy of Drumming: Drums and Percussion Instruments from
  Around the World}.
\newblock Binkey Kok Publications, Diever, Holland, 1997.

\bibitem[Knu98]{knuth-98}
Donald~E. Knuth.
\newblock {\em The Art of Computer Programming}, volume 2, 3rd edition.
\newblock Addison Wesley, Reading, Massachussets, 1998.

\bibitem[Koe70]{koetting-70}
James Koetting.
\newblock Analysis and notation of {W}est {A}frican drum ensemble music.
\newblock {\em Publications of the Institute of Ethnomusicology}, 1(3), 1970.

\bibitem[KR04]{klette-04}
Reinhard Klette and Azriel Rosenfeld.
\newblock Digital straightness - a review.
\newblock {\em Discrete Applied Mathematics}, 139:197--230, 2004.

\bibitem[Lon04]{london-04}
Justin London.
\newblock {\em Hearing in Time: Psychological Aspects of Musical Meter}.
\newblock Oxford University Press, New York, 2004.

\bibitem[Lot02]{lothaire-02}
M.~Lothaire.
\newblock {\em Algebraic Combinatorics on Words}.
\newblock Cambridge University Press, Cambridge, England, 2002.

\bibitem[LP92]{lunnon-92}
W.~F. Lunnon and Peter A.~B. Pleasants.
\newblock Characterization of two-distance sequences.
\newblock {\em Journal of the Australian Mathematical Society (Series A)},
  53:198--218, 1992.

\bibitem[Man85]{manuel-85}
Peter Manuel.
\newblock The anticipated bass in {C}uban popular music.
\newblock {\em Latin American Music Review}, 6(2):249--261, Autumn-Winter 1985.

\bibitem[Mat85]{mathiesen-85}
Thomas~J. Mathiesen.
\newblock Rhythm and meter in ancient {G}reek music.
\newblock {\em Music Theory Spectrum}, 7:159--180, Spring 1985.

\bibitem[Mc{C}98]{mccartin-98}
Brian~J. Mc{C}artin.
\newblock Prelude to musical geometry.
\newblock {\em The College Mathematics Journal}, 29(5):354--370, 1998.

\bibitem[Mon85]{montfort-85}
Matthew Montfort.
\newblock {\em Ancient Traditions--Future Possibilities: Rhythmic Training
  Through the Traditions of Africa, Bali and India}.
\newblock Panoramic Press, Mill Valley, 1985.

\bibitem[Mor96]{morrison-96}
Craig Morrison.
\newblock {\em Go Cat Go: Rockabilly Music and Its Makers}.
\newblock University of Illinois Press, Urbana, 1996.

\bibitem[Mus02]{putumayo-02}
Putumayo~World Music.
\newblock {C}ongo to {C}uba.
\newblock Music CD, 2002.

\bibitem[Ort95]{ortiz-95}
Fernando Ortiz.
\newblock {\em La Clave}.
\newblock Editorial Letras Cubanas, La Habana, Cuba, 1995.

\bibitem[Pal89]{Palasti-1989}
Ilona Pal{\'a}sti.
\newblock A distance problem of {P}. {E}rd{\H{o}}s with some further
  restrictions.
\newblock {\em Discrete Mathematics}, 76(2):155--156, 1989.

\bibitem[PC69]{proca-69}
Vera Proca-Ciortea.
\newblock On rhythm in {R}umanian folk dance.
\newblock {\em Yearbook of the International Folk Music Council}, 1:176--199,
  1969.

\bibitem[Pre83]{pressing-83}
Jeff Pressing.
\newblock Cognitive isomorphisms between pitch and rhythm in world musics:
  {W}est {A}frica, the {B}alkans and {W}estern tonality.
\newblock {\em Studies in Music}, 17:38--61, 1983.

\bibitem[Rah87]{rahn-87}
Jay Rahn.
\newblock Asymmetrical ostinatos in sub-saharan music: time, pitch, and cycles
  reconsidered.
\newblock {\em In Theory Only: Journal of the Michigan Music Theory Society},
  9(7):23--37, 1987.

\bibitem[Rah96]{rahn-96}
Jay Rahn.
\newblock Turning the analysis around: {A}frican-derived rhythms and
  {E}urope-derived music theory.
\newblock {\em Black Music Research Journal}, 16(1):71--89, 1996.

\bibitem[Rap05]{rappaport-05}
David Rappaport.
\newblock Geometry and harmony.
\newblock In {\em Proceedings of BRIDGES: Mathematical Connections in Art,
  Music and Science}, pages 67--72, Banff, Alberta, Canada, 2005.

\bibitem[RD01]{reingold-01}
Edward~M. Reingold and Nachum Dershowitz.
\newblock {\em Calendrical Calculations: The Millenium Edition}.
\newblock Cambridge University Press, Cambridge, England, 2001.

\bibitem[Ric80]{rice-80}
Timothy Rice.
\newblock Aspects of {B}ulgarian musical thought.
\newblock {\em Yearbook of the International Folk Music Council}, 12:43--66,
  1980.

\bibitem[Ric04]{rice-04}
Timothy Rice.
\newblock {\em Music in Bulgaria}.
\newblock Oxford University Press, Oxford, England, 2004.

\bibitem[Rod97]{rodriguez-97}
Olavo~Al\'{e}n Rodr\'{\i}guez.
\newblock {\em Instrumentos de la Musica Folkl\'{o}rico-Popular de Cuba}.
\newblock Centro de Investigaci\'{o}n y Desarrollo de la Musica Cubana, Havana,
  Cuba, 1997.

\bibitem[Ros02]{rosalia-02}
Rene~V. Rosalia.
\newblock {\em Migrated Rhythm: The Tamb\'{u} of Cura\c{c}ao}.
\newblock CaribSeek, 2002.

\bibitem[Sac53]{sachs-53}
Curt Sachs.
\newblock {\em Rhythm and Tempo: A Study in Music History}.
\newblock W. W. Norton, New York, 1953.

\bibitem[Sha94]{shallit-94}
Jeffrey~O. Shallit.
\newblock Pierce expansions and rules for the determination of leap years.
\newblock {\em Fibonacci Quarterly}, 32(5):416--423, 1994.

\bibitem[Sin74]{singer-74}
Alice Singer.
\newblock The metrical structure of {M}acedonian dance.
\newblock {\em Ethnomusicology}, 18(3):379--404, September 1974.

\bibitem[Sol96]{sole-96}
Doug Sole.
\newblock {\em The Soul of Hand Drumming}.
\newblock Mel Bay Productions Inc., Toronto, 1996.

\bibitem[Sta88]{standifer-88}
James~A. Standifer.
\newblock The {T}uareg: their music and dances.
\newblock {\em The Black Perspective in Music}, 16(1):45--62, Spring 1988.

\bibitem[Sti03]{Stillwell}
John Stillwell.
\newblock {\em Elements of Number Theory}.
\newblock Springer, 2003.

\bibitem[Sto05]{stone-05}
Ruth~M. Stone.
\newblock {\em Music in {W}est {A}frica}.
\newblock Oxford University Press, Oxford, England, 2005.

\bibitem[Tou02]{toussaint-02}
Godfried~T. Toussaint.
\newblock A mathematical analysis of {A}frican, {B}razilian, and {C}uban {\em
  clave} rhythms.
\newblock In {\em Proceedings of BRIDGES: Mathematical Connections in Art,
  Music and Science}, pages 157--168, Towson University, Towson, Maryland,
  U.S.A., July 27-29 2002.

\bibitem[Tou03]{toussaint-03}
Godfried~T. Toussaint.
\newblock Classification and phylogenetic analysis of {A}frican ternary rhythm
  timelines.
\newblock In {\em Proceedings of BRIDGES: Mathematical Connections in Art,
  Music and Science}, pages 25--36, Granada, Spain, July 23-27 2003. University
  of Granada.

\bibitem[Tou04a]{Toussaint-2004-CGW}
Godfried~T. Toussaint.
\newblock Computational geometric aspects of musical rhythm.
\newblock In {\em Abstracts of the 14th Annual Fall Workshop on Computational
  Geometry}, pages 47--48, Cambridge, Massachusetts, November 2004.

\bibitem[Tou04b]{toussaint-04b}
Godfried~T. Toussaint.
\newblock A mathematical measure of preference in {A}frican rhythm.
\newblock In {\em Abstracts of Papers Presented to the American Mathematical
  Society}, volume~25, page 248, Phoenix, Arizona, U.S.A., January 7-10 2004.
  American Mathematical Society.

\bibitem[Tou05]{toussaint-05}
Godfried~T. Toussaint.
\newblock Mathematical features for recognizing preference in {S}ub-{S}aharan
  {A}frican traditional rhythm timelines.
\newblock In {\em Proceedings of the 3rd International Conference on Advances
  in Pattern Recognition}, pages 18--27, University of Bath, United Kingdom,
  August 2005.

\bibitem[Tym06]{tymoczko}
Dmitri Tymoczko.
\newblock The geometry of musical chords.
\newblock Manuscript, 2006.
\newblock \url{http://www.music.princeton.edu/~dmitri/voiceleading.pdf}.

\bibitem[Uri93]{uribe-93}
Ed~Uribe.
\newblock {\em The Essence of Brazilian Persussion and Drum Set}.
\newblock CCP/Belwin Inc., Miami, Florida, 1993.

\bibitem[Uri96]{uribe-96}
Ed~Uribe.
\newblock {\em The Essence of Afro-Cuban Persussion and Drum Set}.
\newblock Warner Brothers Publications, Miami, Florida, 1996.

\bibitem[vdL95]{vanderlee-95}
Pedro van~der Lee.
\newblock Zarabanda: esquemas r\'{\i}tmicos de acompa\~{n}amiento en 6/8.
\newblock {\em Latin American Music Review}, 16(2):199--220, Autumn-Winter
  1995.

\bibitem[Wad04]{wade-04}
Bonnie~C. Wade.
\newblock {\em Thinking Musically}.
\newblock Oxford University Press, Oxford, England, 2004.

\bibitem[Win66]{Winograd-1966}
Terry Winograd.
\newblock An analysis of the properties of `deep' scales in a t-tone system.
\newblock Unpublished, May 17 1966.
\newblock Term paper for music theory course at Colorado College.

\bibitem[Woo93]{wooldridge-1993}
Marc Wooldridge.
\newblock {\em Rhythmic Implications of Diatonic Theory: A Study of Scott
  Joplin's Ragtime Piano Works}.
\newblock PhD thesis, State University of New York, Buffalo, 1993.

\bibitem[Wri78]{wright-78}
Owen Wright.
\newblock {\em The Modal System of Arab and Persian Music AD 1250-1300}.
\newblock Oxford University Press, Oxford, England, 1978.

\end{thebibliography}

\newcommand{\etalchar}[1]{$^{#1}$}

\end{document}